\documentclass[a4paper]{aa}

\usepackage{dcolumn}
\usepackage{times}
\usepackage{graphics}
\usepackage{natbib}
\usepackage{supertabular}
\usepackage[figuresright]{rotating}

\newcolumntype{d}[1]{D{.}{.}{#1}}

\bibpunct{(}{)}{;}{a}{}{,}

\begin {document}

\title
 {Far-Infrared Emission from Intracluster Dust in Abell Clusters
  \thanks
   {Based on observations with ISO, an ESA project with instruments
    funded by ESA Member States (especially the PI countries: France,
    Germany, the Netherlands and the United Kingdom) and
    with the participation of ISAS and NASA.}}

\author {M. Stickel\inst{1}
    \and U. Klaas\inst{1}
    \and D. Lemke\inst{1}
    \and K. Mattila\inst{2}}

\institute{Max--Planck--Institut f\"ur Astronomie,
           K\"onigstuhl 17,
           69117 Heidelberg,
           Germany
    \and
           Helsinki University Observatory, P.O.Box 14,
           00014 Helsinki, Finland}

\offprints{M.\ Stickel}

\mail{stickel@mpia-hd.mpg.de}

\date{Received ... ; Accepted ...}

\titlerunning {Intracluster Dust in Abell Clusters}

\authorrunning {M. Stickel et al.}

\abstract
{The ISOPHOT instrument aboard ISO has been used to observe
extended FIR emission of six Abell clusters.  Strip
scanning measurements with crossing position angles centered on the
clusters were carried out at 120\,$\mathrm{\mu m}$ and
180\,$\mathrm{\mu m}$.  The raw profiles of
the $\mathrm{I_{120 \mu m} / I_{180 \mu m}}$ surface brightness ratio
including zodiacal light show a bump
towards Abell 1656 (Coma), dips towards Abell 262 and Abell 2670, and
are without clear structure towards Abell 400, Abell 496, and Abell
4038.  After subtraction of the zodiacal light, the bump towards Abell
1656 is still present, while the dips towards Abell 262 and Abell 2670
are no longer noticable.  This indicates a localized excess of
emitting material outside the Galaxy towards Abell 1656 with
properties different from the galactic foreground cirrus, while the
behavior in Abell 262 and Abell 2670 can be reconciled with galactic
cirrus structures localized on the line-of-sight to these
clusters. The excess of $\approx$ 0.2 MJy/sr seen at 120\,$\mathrm{\mu m}$
towards Abell 1656 (Coma) is interpreted as being due to thermal emission
from intracluster dust distributed in the hot X-ray emitting intracluster
medium. The integrated excess flux within the central region of
10\,$\arcmin$ -- 15\,$\arcmin$ diameter is $\approx$ 2.8 Jy.  Since
the dust temperature is poorly constrained, only a rough estimate of
the associated dust mass of $\mathrm{M_{D} \approx 10^{7}\,M_{\sun}}$
can be derived.  The associated visual extinction is negligible
($\mathrm{A_{V} \ll 0.1 mag}$) and much smaller than claimed from
optical observations. No evidence is found for intracluster dust in
the other five clusters observed.  The absence of any signature for
intracluster dust in five clusters and the rather low inferred dust
mass in Abell 1656 indicates that intracluster dust is likely not
responsible for the excess X-ray absorption seen in cooling flow
clusters. These observations thereby represent a further unsuccessful
attempt in detecting the presumed final stage of the cooling flow
material, in accord with quite a number of previous studies in other
wavelengths regions.  Finally, the observed dimming of the high-redshift
supernovae is unlikely be attributable to extinction caused
by dust in the intracluster or even a presumed intercluster medium.
         }

\maketitle

\section {Introduction}

The analysis of the X-ray and UV emission of the hot
($\mathrm{T_e \approx}$ 1 -- 10 keV)
electron gas trapped in the potential well of galaxy clusters
\citep[e.g.][]{Sarazin86} has revealed an increasingly complex picture
of the intracluster medium (ICM). High resolution X-ray images reveal
the presence of small-scale structures of different temperatures
\citep[e.g.][ and references therein]{Donnellyetal99}.  High central
densities, lower temperatures, and cooling times significantly shorter
than a Hubble time in the centers of many clusters have been
interpreted as evidence for cooling flows
\citep[e.g.][]{Fabian94,Whiteetal97,Allen00}, where the cooling gas
drops out of thermal equilibrium, possibly leading to the
formation of dense clouds and eventually to stars.  X-ray spectra and
image decomposition indicate an extended excess absorption of
soft X-rays with an HI column density of $\mathrm{\approx 10^{21} cm^{-2}}$
above the galactic foreground
\citep[e.g.][]{Whiteetal91,AllenFabian97,Sarazin97,Allen00}. However,
\citet{ArabadjisBregman00} find significantly lower or even no
additional absorbing column densities in a number of other clusters.  A
spatially extended EUV emission of non-thermal origin \citep[][ and
references therein]{Ensslinetal99} above the expected flux from the
hot X-ray gas in at least the Coma and possibly other clusters seems
to be fairly established
\citep{Lieuetal96a,Lieuetal96b,Mittazetal98,Lieuetal99,Bowyeretal99},
although an accompanying low energy X-ray excess emission is uncertain
\citep{Dixonetal96,ArabadjisBregman99}.  Overall, these observations
indicate a striking departure from a nearly homogeneous intracluster
gas towards a spatially inhomogeneous multi-phase ICM
\citep{Bonamenteetal01,Lieuetal00}.

\par
Despite quite some observational effort, neither the final stage of
the cooling intracluster gas nor the material responsible for the
excess absorption has been detected at other
wavelengths \citep[e.g.][ and references therein]
{Sarazin97,Laor97,Koekemoeretal98,Milleretal99,Allen00}.  It
has been suggested that the absorbing material is present as either
very cold molecular clouds \citep{Ferlandetal94} or dense cold clouds
where the shielding allows the formation of dust grains onto which all
volatiles have frozen \citep{Dainesetal94,Fabianetal94}.

\par
The measured line emission of heavy elements
\citep[e.g.][]{Mushotzkyetal96,Fukazawaetal98} clearly indicates that
the ICM contains a processed component, likely the result of galactic
winds, stripping of the interstellar medium (ISM) of cluster members
by ram pressure, or tidally removed ISM during merging with other
galaxy groups. These processes also add dust to the ICM, which in turn
is a possible candidate for the excess X-ray absorber.  The first
direct evidence for dust in the ICM is an observed oxygen K edge in an
X-ray spectrum of the Perseus cluster attributed to neutral material,
most likely dust grains condensed out of the ICM
\citep{ArnaudMushotzky98}.

\par
In equilibrium conditions, the predicted intracluster dust (ICD) temperatures
are rather uncertain with temperatures in the range 10 -- 20 K
\citep{Dweketal90,LoewensteinFabian90,Braineetal95}, but possibly
much higher \citep{Odeaetal94, VoitDonahue95}.
The thermal emission of the ICD is thus to be expected in the FIR.
If not shielded in dense clouds, the high kinetic temperatures destroy
dust grains with a typical size of $\mathrm{\approx 0.1 \mu m}$ by
sputtering over time scales of a few $\mathrm {10^8}$ -- $\mathrm{10^9}$
years \citep{Dweketal90,Tielensetal94,Dweketal96}.  Very big
($\mathrm{> 10 \mu m}$) dust grains introduced to
account for the observed FIR spectra of galaxies
\citep{RowanRobinson92} would have a significantly longer lifetime of
the order a Hubble time.  Possibly, the dust destruction is more
efficient for smaller ($\mathrm{< 0.1 \mu m}$) grains, leaving
preferentially behind the bigger grains, which moreover might have a
flatter extinction curve than those observed in the Milky Way
\citep{Aguirre99b}. Furthermore, the temperatures in the central
cooling flow regions of clusters might generally be lower, which, in
addition to the shielding by clouds, would contribute to an increased
grain lifetime.

\par
If dust is the absorbing material inferred from the X-ray
observations, a significant fraction of the absorbed X-ray radiation
of $\mathrm{\ga 10^{43}\,erg\,s^{-1}}$ is re-emitted in the FIR, and
dust would then constitute a major coolant of the ICM
\citep{Bregman92,Sarazin97}.  The FIR emission from ICD would then
also be relevant for the correct interpretation of measurements of the
Sunyaev-Zeldovich effect in the sub-mm wavelength range
\citep{KhersonskiiVoshchinnikov85,Lamarreetal98}.

\par
Intracluster dust may reveal itself by optical depth effects in
cosmological investigations of distant objects \citep{Shaver87,Masci98}, a
significant fraction of which are viewed through
clusters. Historically, this has been the first attempt to find
evidence for the existence of ICD \citep{Zwicky57,Zwicky62}.
Despite quite some effort to search for the visual extinction of
background quasars and galaxies seen through foreground galaxy
clusters, no common consent has been achieved whether properties
of background sources are actually affected by the presence of dust in
foreground clusters \citep[][ and references therein]
{Girardietal92,Ferguson93,Maoz95}.
\citet{Aguirre99a,Aguirre99b} and \citet{SimonsenHannestad99} have
drawn attention to the possibility that the observed dimming of
distant type Ia supernovae might be caused by intervening
dust rather than resulting from a non-zero cosmological constant.

\par
From a comparison of Ly $\alpha$ and UV continuum fluxes in 10 cooling
flow clusters, \citet{Hu92} found an average excess reddening of
$\mathrm{E_{B-V} \approx 0.19}$ above the galactic foreground
absorption. This was attributed to ICD spread out over a large
fraction of the cluster volume.  Model calculations of the extended
FIR emission from ICD were presented by \citet{Dweketal90} and
\citet{Hu92}, who concluded that the expected FIR flux was just within
reach of the available IRAS data and the then upcoming ISO mission.

\par
The FIR emission as a direct indicator for ICD has been searched for
at 60 $\mathrm{\mu m}$ and 100 $\mathrm{\mu m}$ on IRAS ISSA plates
\citep{Wheelocketal94} by
\citet{Wiseetal93}. Removing the large scale galactic foreground
cirrus emission, evidence for diffuse excess FIR emission of low
statistical significance was found in the direction of several out of
56 clusters.  Sub-millimeter observations of 11 cooling flow clusters
by \citet{AnnisJewitt93} did not detect any emission from dust near
the cluster centers. With the more sensitive sub-millimeter bolometer
array at the JCMT, \citet{Edgeetal99} detected the dust emission of
the central galaxies from 2 out of 5 cooling flow clusters.  However,
the latter two studies were insensitive to extended emission of ICD on
angular scales larger than a few arc minutes.  \citet{Hansenetal00}
used ISOPHOT FIR maps to search for dust coincindent with the central
galaxies of cooling flow clusters, but obtained only inconclusive
results.

\par
Compared to FIR maps, strip-scanning measurements crossing the cluster
provide a means to search a larger fraction of the cluster volume, and
to reach the FIR foreground level far off the cluster centre.
Furthermore, the separation of extended intracluster dust emission
from the foreground zodiacal and galactic cirrus emission can be
accomplished by using observations at two different wavelengths
longward of the IRAS bands.  This combined technique was employed in
a pilot study to search for dust in the intracluster medium of the
Coma cluster (Abell 1656, z = 0.0232) with
ISOPHOT \citep{Lemkeetal96,LemkeKlaas99}, the photometer
instrument aboard ISO \citep{Kessleretal96,Kessler99}.  An enhanced
120\,$\mathrm{\mu m}$ emission within the central region of
$\mathrm{\approx 10\arcmin}$ ($\approx$
0.3 Mpc, $\mathrm{H_{0} = 70\,km\,s^{-1}\,Mpc^{-1}, q_{0} = 0.5}$)
diameter of the cluster was observed \citep{Stickeletal98}. In a
follow-up study, ISOPHOT was used to observe five more galaxy clusters
in a similar way, and the results, together with a re-analysis of the
Coma data, are described in the following.

\begin{table*}
\caption[abelllist]{\label{tab:abelllist} Observed Abell clusters}
\begin{flushleft}
\begin{tabular}{lccd{-1}d{-1}ccr}
\hline
\noalign{\smallskip}
\multicolumn{1}{l}{Cluster}          &
\multicolumn{1}{c}{$\alpha_{2000}$}  &
\multicolumn{1}{c}{$\delta_{2000}$}  &
\multicolumn{1}{c}{l}                &
\multicolumn{1}{c}{b}              &
\multicolumn{1}{c}{Redshift}       &
\multicolumn{1}{c}{Richness-}      &
\multicolumn{1}{c}{$\mathrm{R_{Abell}}$} \\

\multicolumn{1}{c}{}                   &
\multicolumn{1}{c}{}                   &
\multicolumn{1}{c}{}                   &
\multicolumn{1}{c}{[$\degr$]}          &
\multicolumn{1}{c}{[$\degr$]}          &
\multicolumn{1}{c}{}                   &
\multicolumn{1}{c}{Class}              &
\multicolumn{1}{c}{[$\arcmin$]}    \\

\noalign{\smallskip}
\hline\noalign{\smallskip}
\multicolumn{1}{l}{~~~~~~(1)}    &
\multicolumn{1}{c}{(2)}    &
\multicolumn{1}{c}{(3)}    &
\multicolumn{1}{c}{(4)}    &
\multicolumn{1}{c}{(5)}    &
\multicolumn{1}{c}{(6)}    &
\multicolumn{1}{c}{(7)}    &
\multicolumn{1}{c}{(8)}    \\
\hline\noalign{\smallskip}
Abell 262               &  $01^h52^m46.4^s$  &  $+36\degr09\arcmin06\arcsec$  &  136.58  &  -25.09  &  0.0161  &    0  &   110         \\

Abell 400               &  $02^h57^m38.6^s$  &  $+06\degr02\arcmin00\arcsec$  &  170.24  &  -44.94  &  0.0238  &    1  &    75         \\

Abell 496               &  $04^h33^m37.1^s$  &  $-13\degr15\arcmin42\arcsec$  &  209.57  &  -36.48  &  0.0327  &    1  &    56         \\

Abell 1656 (Coma)       &  $12^h59^m35.7^s$  &  $+27\degr57\arcmin38\arcsec$  &   58.16  &  +88.01  &  0.0232  &    2  &    75         \\

Abell 2670              &  $23^h54^m13.7^s$  &  $-10\degr25\arcmin09\arcsec$  &   81.32  &  -68.52  &  0.0761  &    3  &    26         \\

Abell 4038 (Klemola 44) &  $23^h47^m41.9^s$  &  $-28\degr08\arcmin19\arcsec$  &   25.08  &  -75.90  &  0.0283  &    2  &    64         \\

\noalign{\smallskip}
\hline
\end{tabular}
\end{flushleft}

\begin{flushleft}
Notes : \\
- coordinates (Col. 2, 3) of central sky position of ISO scans \\
- redshifts (Col. 6) and richness classes (Col. 7) taken from NED \\
- Abell radii (Col. 8) according to \citet{RudnickOwen77} \\
\end{flushleft}
\end{table*}

\section {Cluster Selection}

In the first intracluster dust observation with ISOPHOT
\citep{Stickeletal98}, the Coma cluster was selected because
it is the cluster where early optical extinction
studies had found evidence for ICD
\citep{Zwicky62,KarachentsevLipovetskii69}, where diffuse optical
emission from intracluster material was already known
\citep{WelchSastry71,Mattila77}, and where a detailed theoretical
study \citep{Dweketal90} showed that the predicted diffuse FIR
emission was within reach for ISOPHOT.

\par
Primary candidates for additional observations were Abell 262 and
Abell 2670, since \citet{Wiseetal93} had found marginal evidence
for extended FIR emission on IRAS ISSA plates. Because the Coma cluster
is known to undergo merging with smaller galaxy groups
\citep{CollessDunn96,Vikhlininetal97,Burnsetal94b} the dust inferred
from the diffuse FIR emission was interpreted as being transferred
rather recently to the ICM by tidal stripping of cluster galaxy ISM
\citep{Stickeletal98}. Extended FIR emission from the ICM thus might
be found preferentially in recent mergers or young clusters, which
makes the testable prediction that old clusters should show weak or no
extended FIR emission.  It was therefore attempted to find more
targets along the sequence of X-ray morphologies outlined by
\citet{JonesForman92} and quantified as a possible dynamical and
evolutionary sequence by \citet{BuoteTsai95,BuoteTsai96}.
Particularly, the supposedly young clusters with irregular and the
supposedly old virialized clusters with circular symmetric X-ray
emission were considered.

\par
Cluster candidates from the list of \citet{BuoteTsai96} were
subject to constraints in the sky visibility by ISO and in cluster
redshift. The latter was necessary because at too large redshifts only
very few non-overlapping sky positions would have been possible to
observe with the 3$\arcmin$ wide ISOPHOT detector inside a
characteristic cluster Abell radius, while the
nearest clusters would have required prohibitively long integration
times to cover a significant fraction of the cluster diameter. Since a
re-observation of the Coma cluster became impossible, Abell 4038 was
chosen as a closely resembling cluster, particularly with respect to
the multipole moments of its X-ray morphology \citep{BuoteTsai96}.

\par
The observed Abell clusters are listed in Tab. \ref{tab:abelllist},
which gives for each cluster (Col. 1) the coordinates, on which the ISOPHOT
observations were centered (Col. 2, 3), its galactic coordinates (Col. 4, 5),
the cluster redshift (Col. 6), the cluster richness class (Col. 7), and the
cluster Abell radius according to \citet{RudnickOwen77} (Col. 8).

\begin{table*}
\caption[logobservation]{\label{tab:logobservation} Observational Details}
\begin{flushleft}
\begin{tabular}{llccclrc}
\hline
\noalign{\smallskip}
Cluster           &  Observing &   ISO         &  \# Steps  & Cluster        &  Position Angle    &  On-target     & Zodiacal Light at                              \\
                  &  Date      &   Revolution  &            & Coverage$^{a}$ &                    &  Time          & 120\,$\mathrm{\mu m}$ /  180\,$\mathrm{\mu m}$ \\
                  &            &               &            &                &  ~~~~[$\degr$]     &  [s]~~~~       &  [MJy/sr]                                      \\
\noalign{\smallskip}
\hline\noalign{\smallskip}
\multicolumn{1}{l}{~~~~~~(1)}  &
\multicolumn{1}{c}{(2)}        &
\multicolumn{1}{c}{(3)}        &
\multicolumn{1}{c}{(4)}        &
\multicolumn{1}{c}{(5)}        &
\multicolumn{1}{l}{~~~~(6)}    &
\multicolumn{1}{r}{(7)~~~~}    &
\multicolumn{1}{c}{(8)}        \\
\hline\noalign{\smallskip}

Abell 262               &  Jan 7, 1998    &   784  &   31   &   0.82   &  45, 135          &  10250  & 2.7 / 1.8 \\
\noalign{\smallskip}

Abell 400               &  Feb 21, 1998   &   829  &   19   &   0.76   &  40, 130          &   6530  & 5.1 / 2.4 \\
\noalign{\smallskip}

Abell 496               &  Mar 12, 1998   &   848  &   17   &   0.91   &  90               &   2950  & 1.9 / 0.9 \\
Abell 496               &  Mar 19, 1998   &   855  &   17   &   0.91   &  180              &   2950  \\
\noalign{\smallskip}

Abell 1656 (Coma)       &  Jul 21, 1996   &   247  &   16   &   0.64   &  36, 82           &   3480  & 2.3 / 1.1 \\
\noalign{\smallskip}

Abell 2670              &  Nov 21, 1997   &   736  &   17   &   1.96   &  19, 84           &   5640  & 3.8 / 1.6 \\
\noalign{\smallskip}

Abell 4038 (Klemola 44) &  Dec 24, 1997   &   769  &   19   &   0.89   &  120              &   3260  & 3.0 / 1.4 \\
Abell 4038 (Klemola 44) &  Dec 25, 1997   &   771  &   19   &   0.89   &   30              &   3260  \\
\noalign{\smallskip}
\hline
\end{tabular}
\end{flushleft}
\begin{flushleft}
Notes : \\
${^a}$ scan length in units of the Abell radius
\end{flushleft}
\end{table*}

\section {Observations}

The observations were carried out with the C200 camera of
ISOPHOT \citep{Lemkeetal96,LemkeKlaas99}, a 2$\times$2 pixel array of stressed
Ge:Ga with a pixel size of $89\farcs$4. For each cluster, linear scans
across the cluster center at two different position angles were
obtained, each of which was observed with both the C\_120 and
C\_180 filters (reference wavelength 120\,$\mathrm{\mu m}$ and
180\,$\mathrm{\mu m}$, equivalent widths 47\,$\mathrm{\mu m}$ and
72\,$\mathrm{\mu m}$, respectively) at the same sky positions.
The positional offset between subsequent sky measurements was
3$\arcmin$, which resulted in almost no detector overlap.  The
integration time at each sky position was $\approx$ 50\,s for Abell 1656
and $\approx$ 80\,s for the other clusters, during which 6 -- 16
integration ramps with 127 non-destructive readouts were obtained.

\par
Details of the observations are listed in Tab. \ref{tab:logobservation},
which gives for each cluster (Col. 1) the date (Col. 2) and ISO
revolution number of the observation (Col. 3), the number of sky
positions (Col. 4) and the fraction of the Abell radius covered by
the scan (Col. 5) at each position angle (Col. 6) together with
the total on-target integration time (Col. 7).  Finally, Col. 8
gives the zodiacal light contribution for 120\,$\mathrm{\mu m}$
and 180\,$\mathrm{\mu m}$ for the observing date in Col. 2 taken
from \citet{Kelsalletal98} for all clusters except Abell 2670, which
in turn was taken from the yearly averaged zodiacal light map
given by \citet{Leinertetal98}.

\begin{table*}
\caption[physicalproperties]{\label{tab:physicalproperties} Physical Properties of Observed Clusters}
\begin{flushleft}
\begin{tabular}{lcrrrrrrr}
\hline
\noalign{\smallskip}
Cluster      &   X-Ray        &  $\mathrm{N_HI}$               &   Temperature  &  Cooling Flow                     &   GasMass                         &  $\mathrm{L_{XBol}}$                      \\
             &   Morphology   &  [$\mathrm{10^{20} cm^{-2}}$]  &    [keV]       &  [$\mathrm{M_{\sun}\,yr^{-1}}$]   &   [$\mathrm{10^{12} M_{\sun}}$]   &  [$\mathrm{10^{44} erg\,s^{-1}}$]         \\
\noalign{\smallskip}
\hline\noalign{\smallskip}
\multicolumn{1}{c}{(1)}    &
\multicolumn{1}{c}{(2)}    &
\multicolumn{1}{c}{(3)}    &
\multicolumn{1}{c}{(4)}    &
\multicolumn{1}{c}{(5)}    &
\multicolumn{1}{c}{(6)}    &
\multicolumn{1}{c}{(7)}   \\
\hline\noalign{\smallskip}
Abell 262         &  elliptical ~[4]       &   5.3 ~[1]     &   2.4 ~[1]      &    27 ~[3]      &     2.5 ~[3]   &   0.9 ~[1]       \\
                  &                        &  10.5 ~[4]     &   2.5 ~[2]      &     9 ~[2]      &     6.5 ~[2]   &   0.5 ~[2]       \\
                  &                        &                &   1.4 ~[4]      &    47 ~[7]      &                &   0.4 ~[4]       \\
\noalign{\smallskip}
Abell 400         &  irregular ~[6,8]      &   8.8 ~[1]     &   2.5 ~[1]      &     0 ~[2]      &    11.1 ~[2]   &   0.7 ~[1]       \\
                  &                        &                &   2.2 ~[2]      &     0 ~[10]     &    15.0 ~[6]   &   0.6 ~[2]       \\
\noalign{\smallskip}
Abell 496         &  single                &   4.4 ~[1]     &   3.9 ~[1]      &   112 ~[7]      &     2.7 ~[2]   &   5.8 ~[1]       \\
                  &  symmetric  ~[8]       &  12.0 ~[5]     &   4.8 ~[2]      &   138 ~[2]      &    74.1 ~[2]   &   7.9 ~[2]       \\
                  &                        &                &   3.3 ~[5]      &    95 ~[3]      &                &                  \\
\noalign{\smallskip}
Abell 1656        &    elliptical ~[11]    &   0.9 ~[1]     &   8.3 ~[1]      &     0 ~[3]      &    27.8 ~[3]   &  12.5 ~[1]       \\
(Coma)            &                        &                &   7.3 ~[2]      &     0 ~[2]      &    75.6 ~[2]   &   9.1 ~[2]       \\
                  &                        &                &                 &                 &                &                  \\
\noalign{\smallskip}
Abell 2670        &    double ~[8,9]       &   2.7 ~[1]     &   3.9 ~[1]      &    41 ~[2]      &    40.7 ~[2]   &   3.8 ~[1]       \\
                  &                        &   3.6 ~[9]     &   3.7 ~[2]      &     0 ~[9]      &                &   3.5 ~[2]       \\
                  &                        &                &   4.2 ~[9]      &    18 ~[12]     &                &                  \\
\noalign{\smallskip}
Abell 4038        &    elliptical ~[8]     &   1.5 ~[1]     &   3.3 ~[1]      &    87 ~[3]      &     4.3 ~[3]   &   2.2 ~[1]       \\
(Klemola 44)      &                        &                &                 &                 &                &                  \\
\noalign{\smallskip}
\hline
\end{tabular}
\end{flushleft}
\begin{flushleft}
References : \\
{[1]} David et al. (1993);~~
{[2]} White et al. (1997);~~
{[3]} Peres et al. (1998);~~
{[4]} David et al. (1996);~~
{[5]} MacKenzie et al. (1996);~~
{[6]} Beers et al. (1992);~~
{[7]} Edge et al. (1992);~~
{[8]} Buote \& Tsai (1996);~~
{[9]} Hobbs \& Willmore (1997);~~
{[10]} Allen et al. (1997);~~
{[11]} Vikhlinin et al. (1997);~~
{[12]} Wise et al. (1993)
\end{flushleft}
\end{table*}

\par
Physical properties of the observed clusters are collected in Tab.
\ref{tab:physicalproperties}, which gives for each cluster (Col. 1)
the overall X-ray morphology (Col. 2), the galactic foreground column
density (Col. 3), the X-ray temperatures (Col. 4), the estimated
cooling flow rate (Col. 5), the derived gas mass (Col. 6), and the
X-ray luminosity (Col. 7). Where possible, several entries for each
quantity together with the literature reference have been listed to
give an indication for the spread in the determination of these
values. The six observed clusters cover the range from relaxed
elliptical (Abell 262) to irregular (Abell 400) X-ray morphologies,
from cool (Abell 262, Abell 400) to hot (Abell 1656) X-ray
temperatures, from no (Abell 400, Abell 1656) to a significant (Abell
496) cooling flow rate, and about an order of magnitude in galactic
$\mathrm {N_{HI}}$ column density.

\section {Data Reduction}

\par
To push the detection of the expected weak signals at these long
wavelengths to the limits, and to get confidence in detected features,
several different methods for the signal derivation were used in
parallel. Already in the case of the Coma cluster
\citep{Stickeletal98}, the pairwise differences of consecutive ramp
read-outs rather than signals from the full ramp fitting have proven
to provide the most robust signals, particularly if only a few ramps
at each sky position are available.
To get rid of pairwise readout differences affected by cosmic ray
hits, the robust outlier-insensitive myriad estimator was computed and
15\% of the most deviant signals as measured by the absolute deviation
were cut off. This outlier removal is similar to a median absolute
deviation trimming, but instead of the initial median, the sample
myriad is used to determine the outliers.  The sample myriad value in
turn is a robust estimator of the mode (most common value) of a
distribution but does not require binning of the actual data set, and
is easily computed by minimizing a particular cost function with a
tuning constant set to a small value \citep[for details see][]{KalluriArce98}.

\par
Separate sets of final signals for each sky position were derived from
this tail-trimmed pairwise distribution by computing the mean, median,
myriad \citep{KalluriArce98}, and the an\-nealing M
estimator \citep{Li96}. Additionally, the cumulative trimmed distribution
was fitted with the analytical function
$\mathrm {y = c_{0}(1 + exp((x-c_{3})/c_{1}))^{c_{2}})}$,
the first derivative of which resembles very closely a gaussian.  In
this case, the final signal value was derived from the maximum of the
first derivative of the fitting function.

\par
Signals from the much smaller distribution of ramp slopes at each sky
position, obtained by fitting first order polynomials to the
deglitched readouts of each integration ramp, were derived for
comparison.  The initial ramp-fitting step was done within the ISOPHOT
Interactive Analysis
PIA\footnote{The ISOPHOT data presented in this paper were reduced using
             PIA, which is a joint development by the ESA Astrophysics Division
             and the ISOPHOT Consortium. The ISOPHOT Consortium is led by the
             Max-Planck-Institut f\"ur Astronomie, Heidelberg.}
Version 9.0 \citep{Gabrieletal97}.
Since the number of ramps for one sky position was
rather small, a subsequent efficient rejection of disturbed ramp
slopes was not possible. Therefore, only the robust myriad,
the annealing M estimator, and the standard PIA averaging were applied to
derive the final signals.

\par
Eventually, the signals derived by all different methods were
corrected for signal dependence on ramp integration times to be
consistent with calibration observations \citep{Laureijsetal00},
dark-current subtracted, and finally flux calibrated with PIA Version
9.0 / Cal~G Version 6.0.  For the conversion to an absolute flux
level, measurements of the ISOPHOT internal Fine Calibration Source
(FCS) obtained at the beginning and end of each scan in each filter
were used.  The signals from the two accompanying FCS measurements
were linearly interpolated in time to calibrate each sky position of
each scan in each filter separately.  Because the ISOPHOT calibration
has been updated since the publication of the Coma results
\citep{Stickeletal98} it was re-analyzed to get a homogeneous set of
data.

\par
A comparison of the various signal derivation methods
showed that the most robust results as judged by the scatter in the
scan signals and $\mathrm{I_{120 \mu m} / I_{180 \mu m}}$ surface
brightness ratios were obtained with the trimmed pairwise distribution
using the myriad value as central estimator, while the median and mean
values of the trimmed distribution showed a larger scatter.
Therefore, only the calibrated data derived from the pairwise myriad method
will be considered further. However, the gross features in the
$\mathrm{I_{120 \mu m} / I_{180 \mu m}}$ surface brightness ratios
described below were present with decreasing confidence in the
$\mathrm{I_{120 \mu m} / I_{180 \mu m}}$ profile of all different
methods used.

\begin{figure}
\centering
\resizebox{\hsize}{!}{\includegraphics{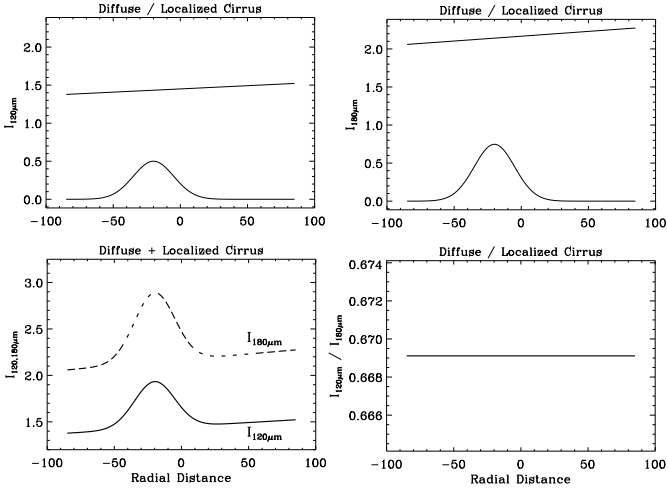}}
\caption[h3052f1.ps]{\label{fig:ModelPanel1}
          One-dimensional model of a FIR scan measurement
          containing cirrus only.
          A diffuse (straight line) and a localized cirrus component
          (gaussian) at
          120\,$\mathrm{\mu m}$ (top left) and 180\,$\mathrm{\mu m}$
          (top right) have similar temperatures, hence the summed
          intensities at
          120\,$\mathrm{\mu m}$ (bottom left, continuous) and
          180\,$\mathrm{\mu m}$ (bottom left, dashed)
          have a constant ratio (bottom right).}
\end{figure}

\begin{figure}
\centering
\resizebox{\hsize}{!}{\includegraphics{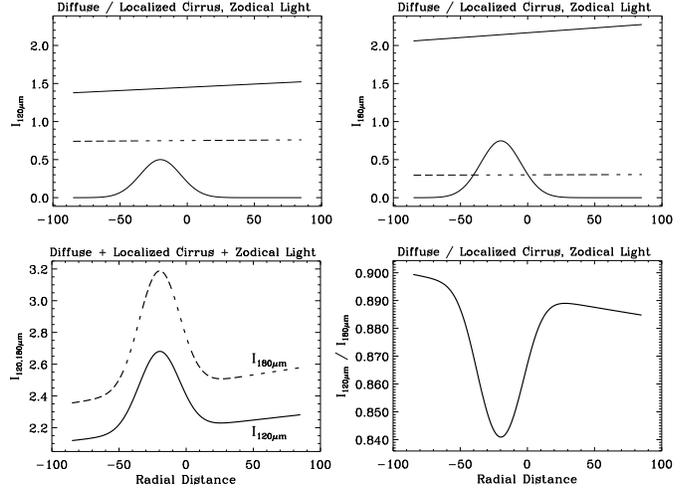}}
\caption[h3052f2.ps]{\label{fig:ModelPanel2}
          One-dimensional model of a FIR scan measurement
          with cirrus and zodiacal light.
          In addition to the diffuse (continuous straight line) and a
          localized cirrus component
          (gaussian), a zodiacal light component (dashed line)
          with a much higher temperature is present at
          120\,$\mathrm{\mu m}$ (top left) and 180\,$\mathrm{\mu m}$
          (top right).
          Although the summed intensities
          (bottom left) look similar to the case without zodiacal light,
          their ratio (bottom right) does reflect the profile of the
           localized cirrus structure as a dip.}
\end{figure}

\section {Data Analysis}

To search for systematic trends in the FIR color profiles across the
clusters, the $\mathrm{I_{120 \mu m} / I_{180 \mu m}}$ ratios for each
position angle were computed for each detector pixel separately and
subsequently averaged.  If necessary, at most a single outlier at each
sky position was rejected.  To further reduce the scatter in the
$\mathrm{I_{120 \mu m} / I_{180 \mu m}}$ ratios, peaks from point
sources were removed by linearly interpolating between the two
adjacent sky positions.  The overall flux ratio distribution was
derived by averaging the individual ratios of all pixels and both PAs,
applying a global rescaling or outlier rejection if necessary.

\par
Care has to be taken if in addition to the diffuse large scale FIR
emission from the galaxy a localized cirrus structure having a similar
temperature is crossed. Even the small contribution from the at FIR
wavelengths weak zodiacal light will break the degeneracy of the
diffuse and localized cirrus component in
the $\mathrm{I_{120 \mu m} / I_{180 \mu m}}$ ratio and
mimic an additional extended FIR emitting component.

\par
Specifically, if the 120\,$\mathrm{\mu m}$ and 180\,$\mathrm{\mu m}$
surface brightnesses of the diffuse cirrus component are related
by $\mathrm{I_{120}^{dif} = R^{cirr}\,\times\,I_{180}^{dif}}$,
and the brightnesses of a localized cirrus structure with a similar
temperature are given by
$\mathrm{I_{120}^{loc} = R^{cirr}\,\times\,I_{180}^{loc}}$,
the  $\mathrm{I_{120 \mu m} / I_{180 \mu m}}$ ratio  $\mathrm{R^{cirr}}$
of galactic cirrus alone is given by

\begin{equation}
\frac{I^{dif}_{120} + I^{loc}_{120}}{I^{dif}_{180} + I^{loc}_{180}} = R^{cirr}.
\label{equ:relation1}
\end{equation}

\par
For similar temperatures of both components, this ratio is nearly
constant, and almost independent of the brightness profiles of both
cirrus components. For an additional zodiacal light component
the brightnesses at both wavelengths are related by
$\mathrm{I_{120}^{zodi} = R^{zodi}\,\times\,I_{180}^{zodi}}$,
where $R^{zodi} \not= R^{cirr}$ because of the different temperatures
of galactic cirrus and zodiacal light.

\par
The total observed $\mathrm{I_{120 \mu m} / I_{180 \mu m}}$
ratio is then given by
\begin{eqnarray}
\lefteqn{\frac{{I^{dif}_{120}} + {I^{loc}_{120}} + {I^{zodi}_{120}}}
               {I^{dif}_{180} + I^{loc}_{180} + I^{zodi}_{180}}
              = R^{cirr} \,+ } \hspace{5em} \nonumber \\
& & (R^{zodi} - R^{cirr})\,\times\,\frac{I^{zodi}_{180}}{I^{dif}_{180} + I^{loc}_{180} + I^{zodi}_{180}}~.
\label{equ:relation2}
\end{eqnarray}

\par
Although the brightnesses of the diffuse cirrus component and the
zodiacal light vary only slightly across the cluster, the overall
ratio is not constant due to the varying brightness of the localized
cirrus structure $\mathrm{I_{180 \mu m}^{loc}}$, and directly reflects
the brightness profile of this component.

\par
This behavior can be exemplified by a one-dimensional model, where
the diffuse cirrus component $\mathrm{I^{dif}}$ might be
approximated by a slanted line with a
constant ratio $\mathrm{I^{dif}_{120}}$ / $\mathrm{I^{dif}_{180}}$,
while the localized cirrus structure might be modeled for simplicity
by a gaussian with a constant $\mathrm{I^{loc}_{120}}$ /
$\mathrm{I^{loc}_{180}}$ ratio equal to that of the diffuse
component. The zodiacal light component can also be described by a
slanted line to account for a gradient in the zodiacal light
distribution along the scan path.

While the model profile of $R^{cirr}$ (Eq. \ref{equ:relation1}) is
constant (Fig. \ref{fig:ModelPanel1}), the zodiacal light component in $R^{cirr+zodi}$
(Eq. \ref{equ:relation2}) makes the presence of a localized cirrus
component obvious, leading to a characteristic change in
the $\mathrm{I_{120 \mu m} / I_{180 \mu m}}$ ratio along the model scan,
which actually reflects the profile of the localized cirrus component
(Fig. \ref{fig:ModelPanel2}).

On the other hand, if no localized cirrus structure
but instead an ICD component described
by $\mathrm{I_{120 \mu m}^{ICD} = R^{ICD}\,\times\,I_{180 \mu m}^{IDC}}$
is present with a temperature different from the galactic
cirrus, i.e. $R^{ICD} \not= R^{cirr}$, then the ratio
\begin{eqnarray}
\lefteqn{\frac{I^{dif}_{120} + I^{ICD}_{120}}
              {I^{dif}_{180} + I^{ICD}_{180}} = R^{cirr} + } \hspace{5em} \nonumber \\
& & (R^{ICD} - R^{cirr})\,\times\,\frac{I^{ICD}_{180}}{I^{dif}_{180} + I^{ICD}_{180}}
\label{equ:relation3}
\end{eqnarray}

\noindent
is already varying across the cluster due to the non-constant ICD
component, reflecting the ICD brightness profile.  Adding the zodiacal
light only changes the level of variation in the observed
overall $\mathrm{I_{120 \mu m} / I_{180 \mu m}}$ ratio, but not the
presence of a dip or bump originating in the ICD component.

\par
This behavior provides the possibility to distinguish in the zodiacal
light subtracted $\mathrm{I_{120 \mu m} / I_{180 \mu m}}$ ratio an
additional ICD component with a surface brightness varying on scales
of a clusters diameter from a localized cirrus component. If a dip or
bump is still present in the $\mathrm{I_{120 \mu m} / I_{180 \mu m}}$
ratio after the subtraction of the zodiacal light component, it can be
attributed to an additional FIR component outside the Galaxy and most
likely identified with the thermal FIR emission of ICD.  On the other
hand, a dip or bump which disappears in
the $\mathrm{I_{120 \mu m} / I_{180 \mu m}}$
ratio with the subtraction of the zodiacal light component is most
likely due to a localized cirrus structure. Only if the ICD
properties (temperature, grain composition, etc.)
closely resemble that of galactic cirrus, a characteristic
change of the $\mathrm{I_{120 \mu m} / I_{180 \mu m}}$ ratio from ICD
will also disappear with the subtraction of the zodiacal light
component, and there is then no possibility to separate the two
components with the measurements at two FIR wavelengths.

\section {Results}

\subsection {Abell 262}

\begin{figure}
  \centering
  \resizebox{\hsize}{!}{\includegraphics{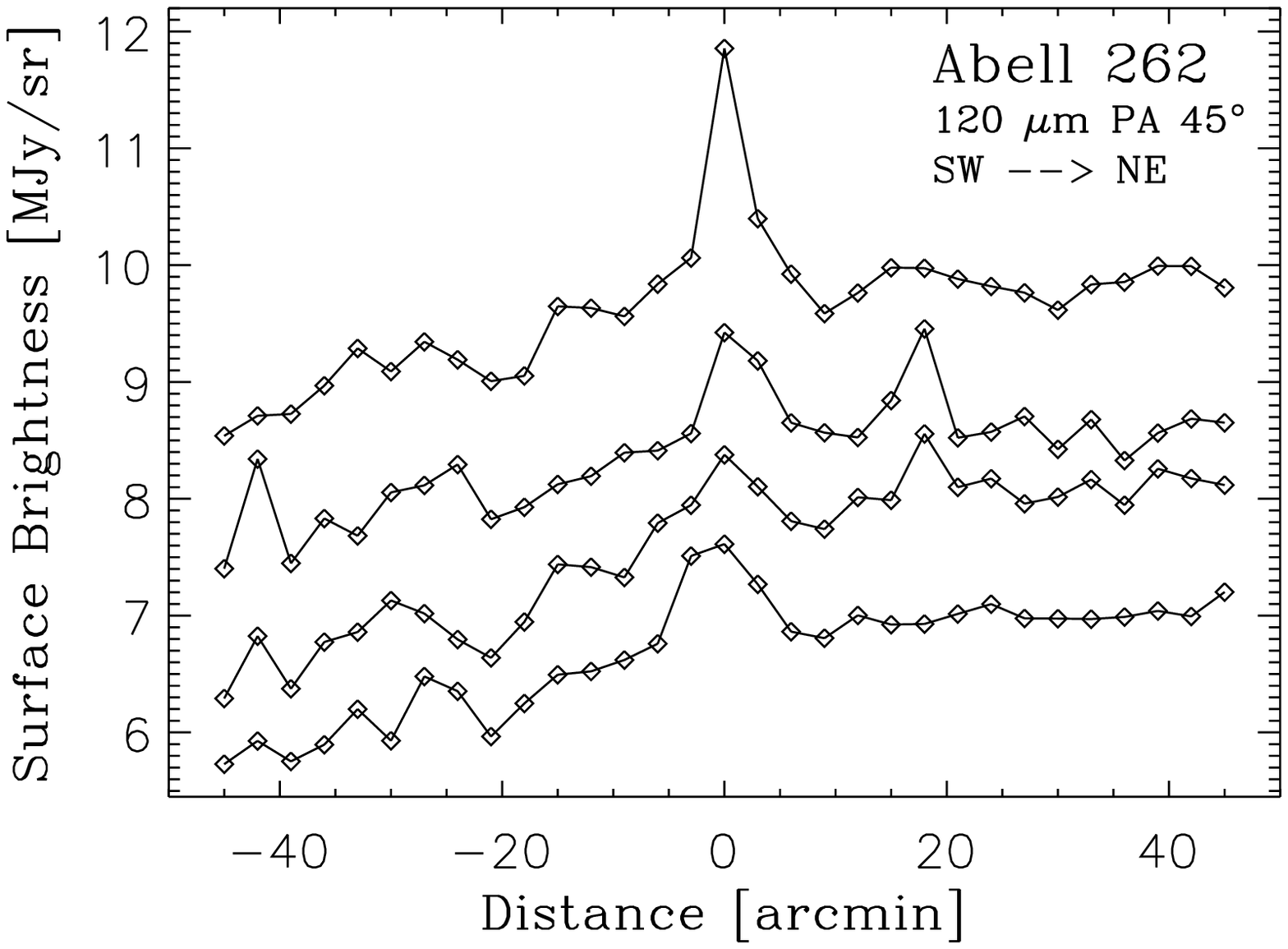}}
  \resizebox{\hsize}{!}{\includegraphics{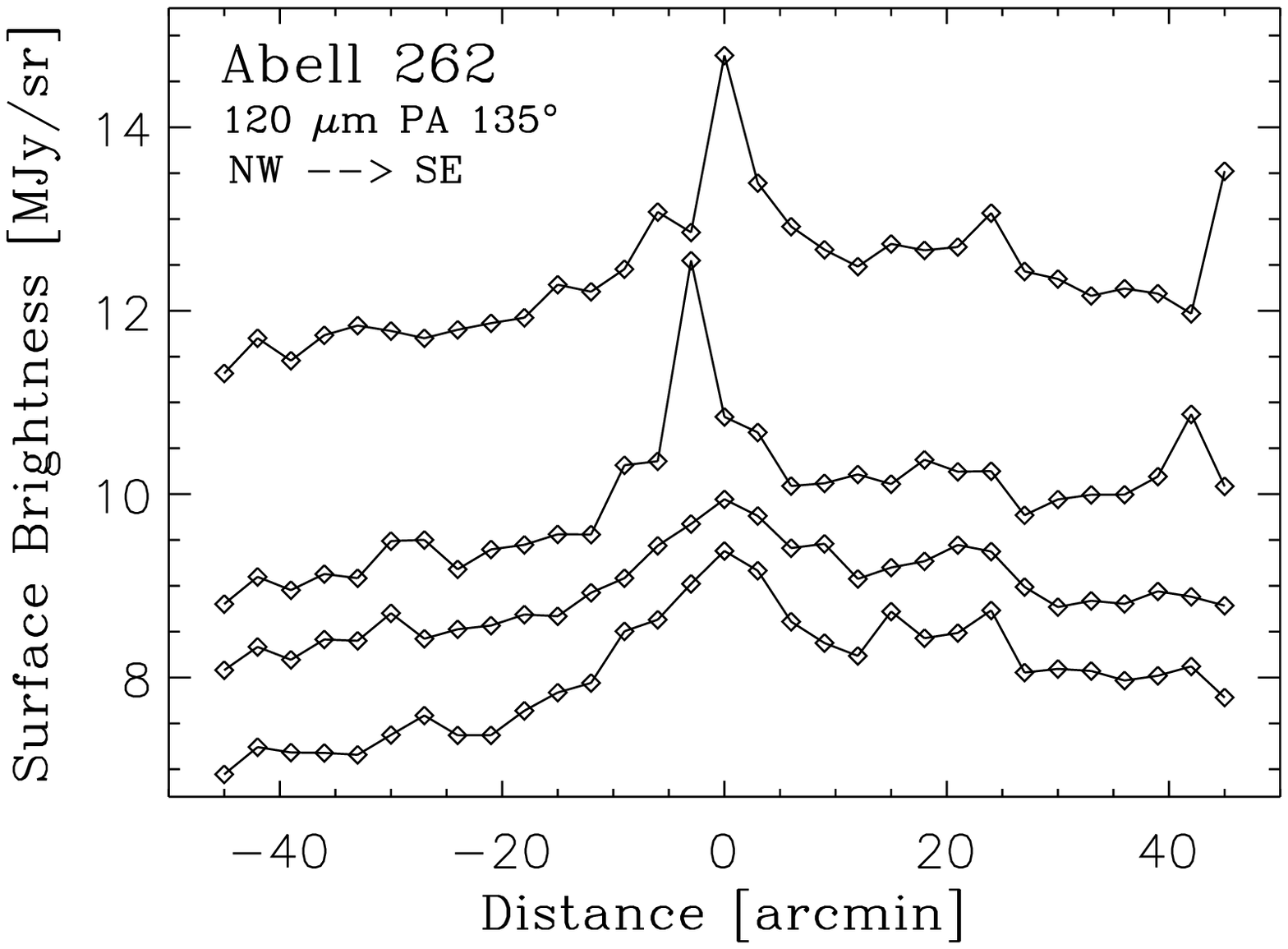}}
  \caption[]{\label{fig:Abell262measured}
           The observed brightness distributions of the four C200 detector
           pixels at 120 $\mathrm{\mu m}$ for Abell 262 along PA
           45$\degr$ (top) and PA 135$\degr$ (bottom). South lies to the left
           at negative distances, north to the right.  The brightness
           level is correct only for the lowest data stream. For
           clarity, the other three pixel data streams are offset
           arbitrarily. The 180\,$\mathrm{\mu m}$ brightness
           distributions (not shown) are quite similar.}
\end{figure}

\begin{figure}
\centering
\resizebox{\hsize}{!}{\includegraphics{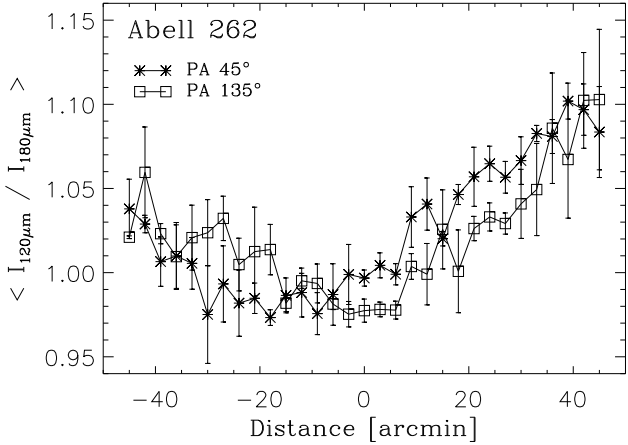}}
\caption[h3052f4.ps]{\label{fig:Abell262RatiosWZodi}
         The raw surface brightness ratios
         $\mathrm{I_{120 \mu m} / I_{180 \mu m}}$, averaged over
         all four detector pixels, along
         PA 45$\degr$ (asterisks) and PA 135$\degr$ (squares) as a
         function of distance from the cluster center of Abell 262.
         South lies to the left at negative distances, north to the
         right. Both PAs show a marked south-north asymmetry and a
         broad depression closely aligned with the cluster center.}
\end{figure}

\begin{figure}
\centering
\resizebox{\hsize}{!}{\includegraphics{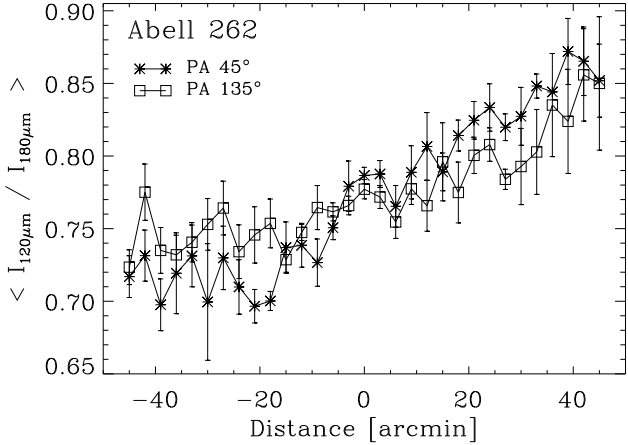}}
\caption[h3052f5.ps]{\label{fig:Abell262Ratios}
         The surface brightness ratios $\mathrm{I_{120 \mu m} / I_{180\mu m}}$
         after subtraction of the Zodiacal light,
         averaged over all four detector pixels, along
         PA 45$\degr$ (asterisks) and PA 135$\degr$ (squares) as a
         function of distance from the cluster center of Abell 262.
         South lies to the left at negative distances, north to the
         right. }
\end{figure}

\begin{figure}
\centering
\resizebox{\hsize}{!}{\includegraphics{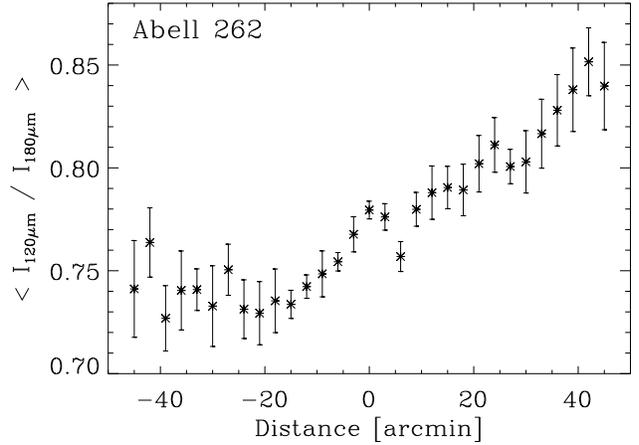}}
\caption[h3052f6.ps]{\label{fig:Abell262Overall}
         The overall zodiacal-light subtracted surface brightness ratio
         $\mathrm{I_{120 \mu m} / I_{180 \mu m}}$ for Abell 262,
         averaged over both position angles and all
         detector pixels.}
\end{figure}

\par
Abell 262 is a spiral rich cluster with an E/D type central galaxy (NGC 708)
showing dust lanes and diffuse gas with line emission.  The
cluster X-ray temperature is rather low, and there is independent
evidence for a central cooling flow
(Tab. \ref{tab:physicalproperties}). \citet{GiovanelliHaynes85} found
that a significant fraction of the galaxies inside one Abell radius
are deficient in HI, while \citet{BravoAlfaroetal97} have shown that a
number of these galaxies have an asymmetric HI distribution. Both
results are most likely the results of the interaction of the cluster
galaxies with the ICM.

\par
The 100 $\mathrm{\mu m}$ IRAS HIRES image
shows an elongated cloud-like patch of FIR
emission, roughly oriented at a PA $\approx 50\degr$, overlaid by
filamentary structures.  One ISOPHOT scan was
executed parallel to this general elongation, while the other one was
perpendicular.  The extended 100 $\mathrm{\mu m}$ emission region lies
at the edge of a much larger cirrus structure, and \citet{Davidetal96}
argued that the excess X-ray absorber required by the ROSAT PSPC
observations is actually galactic cirrus rather than an absorber
physically associated with the cluster. The inferred column density of
galactic HI is about twice the value usually used for the analysis of
X-ray data \citep{Starketal92}, indicating significant fluctuations in the
galactic HI column density on angular scales $< 2\degr$.

\par
However, from emission line ratios \citet{Hu92} found evidence for an
excess reddening of $\mathrm {E_{B-V} \approx}$ 0.20 mag, indicating a
cluster internal absorber containing dust.  \citet{Wiseetal93}
attempted to subtract the galactic foreground cirrus from IRAS ISSA plates
and found evidence for residual diffuse extended emission at 60
$\mathrm{\mu m}$ and 100 $\mathrm{\mu m}$, which was attributed
to intracluster dust, too.

\par
Both ISOPHOT scans show for both position angles the broad bump of
extended FIR emission, and confined to one sky position, the presence
of an unresolved central point source
(Fig. \ref{fig:Abell262measured}).  It can not be decided whether this
FIR point source is due to the central cD galaxy, because a spiral galaxy
is projected onto its outskirts \citep{Fantietal82}.  This confusion
was already noted by \citet{Bregmanetal90}.  Several other
compact sources appear only in the data streams of individual pixels,
indicating non-central crossings.

\par
The raw $\mathrm{I_{120 \mu m} / I_{180 \mu m}}$ surface brightness
ratios including zodiacal light for both position angles, separately
averaged over the four detector pixels, agree remarkably well
(Fig. \ref{fig:Abell262RatiosWZodi}), even in the marked north - south
asymmetry, and show a very broad minimum somewhat off-center from the
nominal cluster centre position.  Remarkably, the broad depression
completely disappears in the zodiacal-light subtracted surface
brightness ratios along both PAs (Fig. \ref{fig:Abell262Ratios}),
leaving only an almost linearly increasing
overall $\mathrm{I_{120 \mu m} / I_{180 \mu m}}$ ratio
(Fig. \ref{fig:Abell262Overall}).

\subsection {Abell 400}

\begin{figure}
  \centering
  \resizebox{\hsize}{!}{\includegraphics{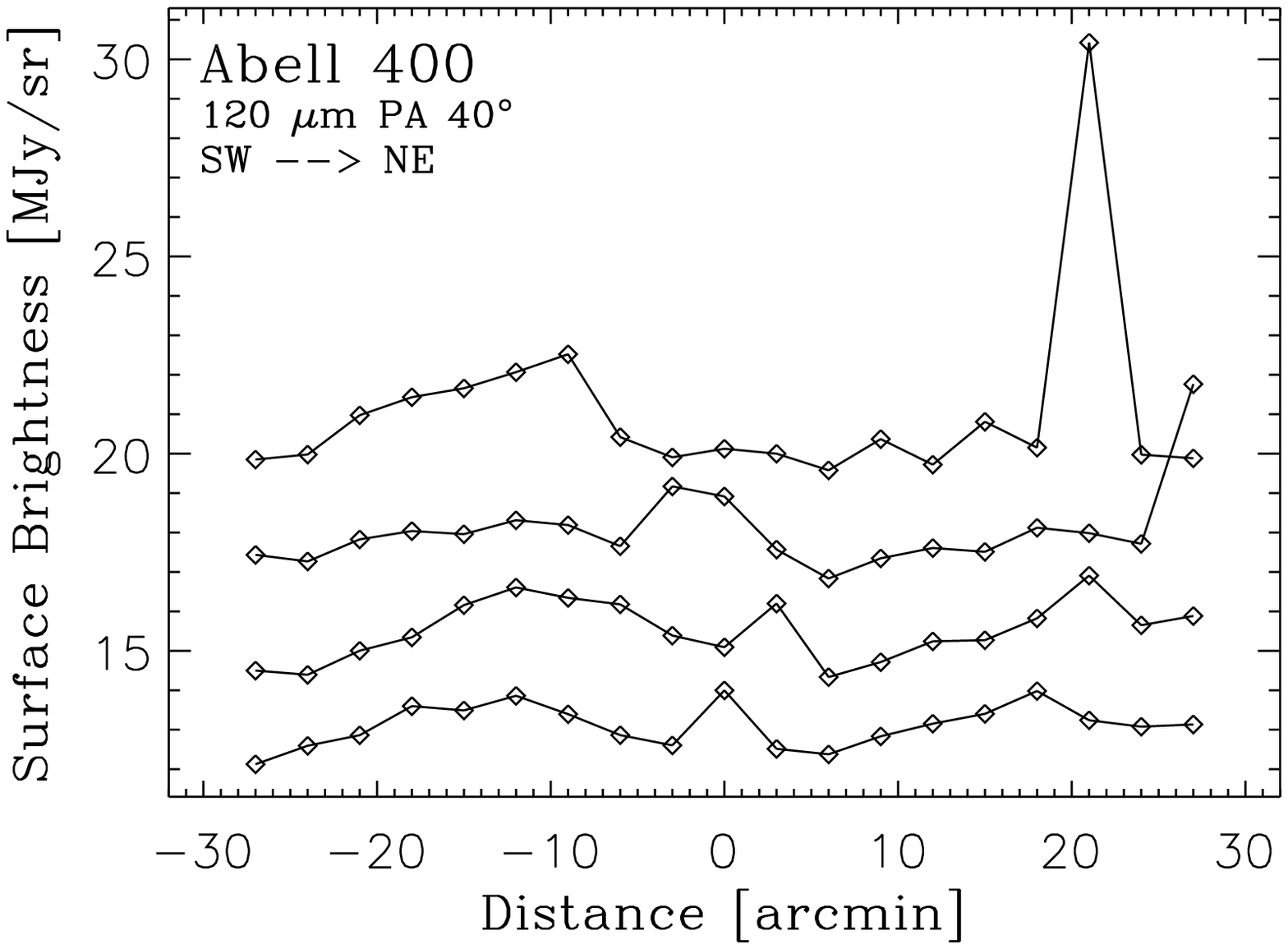}}
  \resizebox{\hsize}{!}{\includegraphics{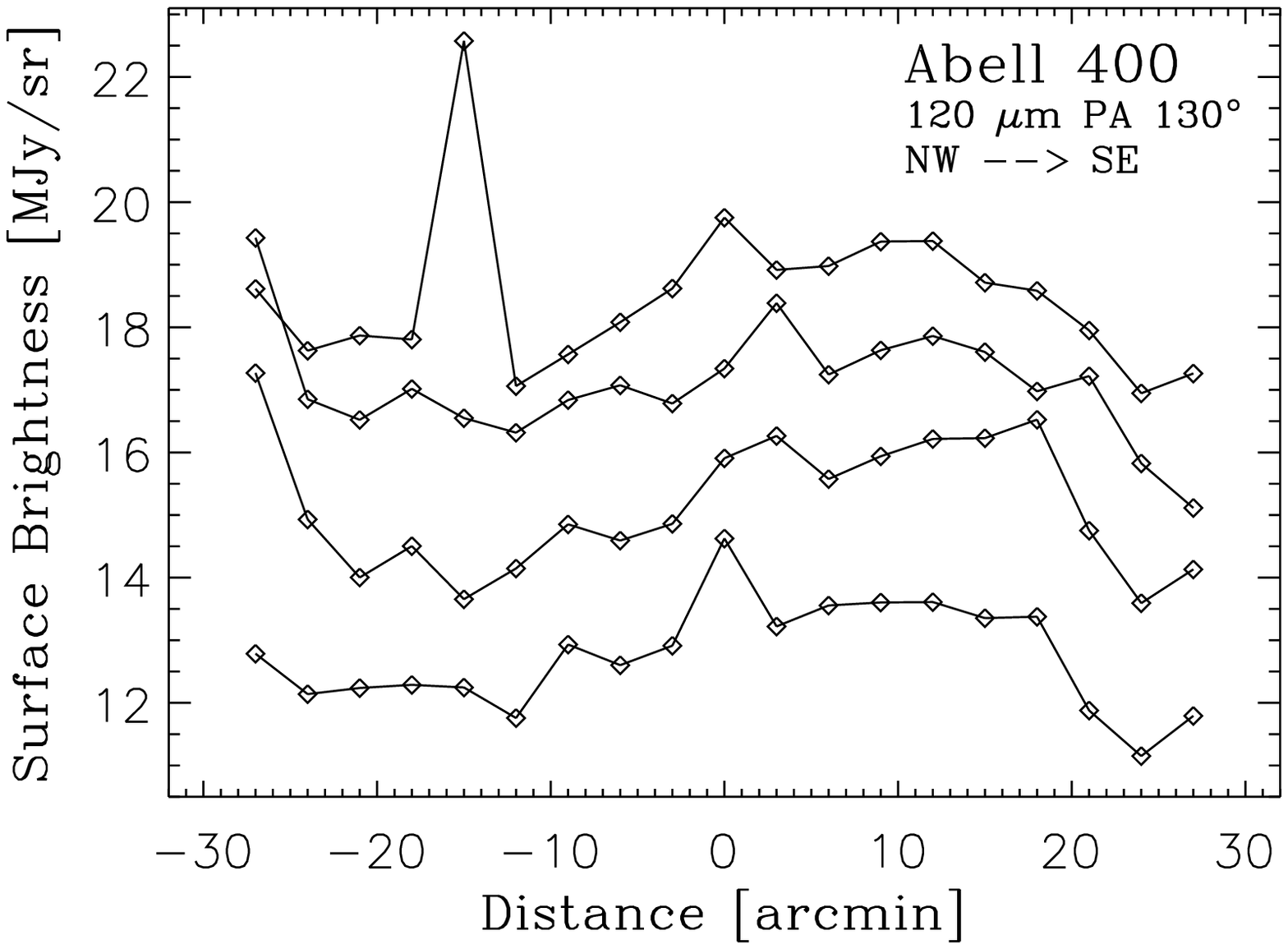}}
  \caption[]{\label{fig:Abell400measured}
           The observed brightness distributions of the four C200 detector
           pixels at 120\,$\mathrm{\mu m}$ for Abell 400 along PA
           40$\degr$ (top) and PA 130$\degr$ (bottom). South lies to the left
           at negative distances, north to the right.  The brightness
           level is correct only for the lowest data stream. For
           clarity, the other three pixel data streams are offset
           arbitrarily. The 180\,$\mathrm{\mu m}$ brightness
           distributions (not shown) are quite similar.}
\end{figure}

\begin{figure}
\centering
\resizebox{\hsize}{!}{\includegraphics{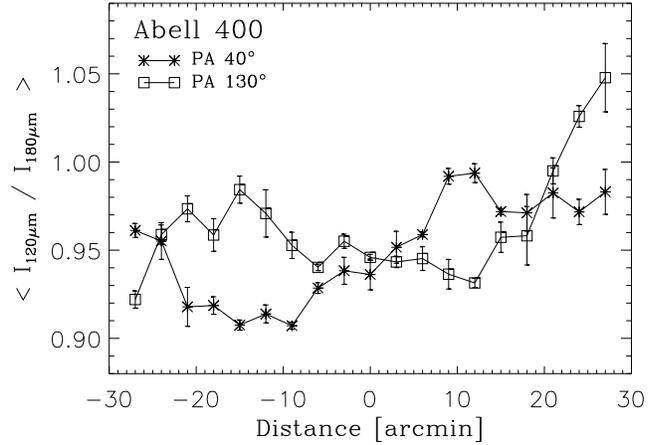}}
\caption[h3052f8.ps]{\label{fig:Abell400RatiosWZodi}
         The raw surface brightness ratios $\mathrm{I_{120 \mu m} / I_{180 \mu m}}$
         averaged over all four detector pixels,
         along PA 40$\degr$ (asterisks) and PA 130$\degr$ (squares) as
         a function of distance from the center of Abell 400.  The
         profiles along the two PAs are rather dissimilar
         with structures on angular scales of 10$\arcmin$ --  20$\arcmin$.
         South lies to the left at negative distances, north to
         the right.}
\end{figure}

\begin{figure}
\centering
\resizebox{\hsize}{!}{\includegraphics{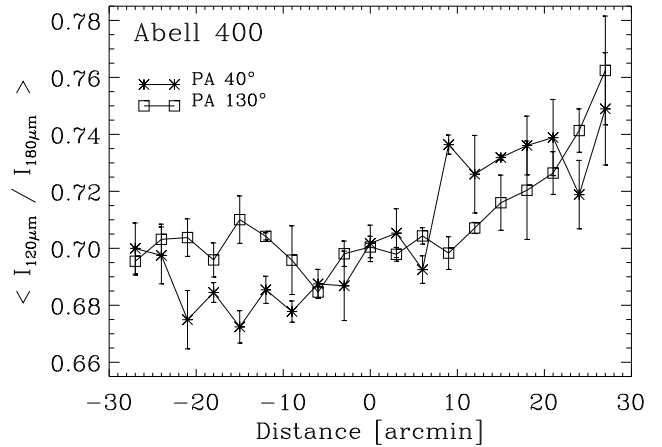}}
\caption[h3052f9.ps]{\label{fig:Abell400Ratios}
         The surface brightness ratios $\mathrm{I_{120 \mu m} / I_{180 \mu m}}$
         after subtraction of the Zodiacal light,
         averaged over all four detector pixels,
         along PA 40$\degr$ (asterisks) and PA 130$\degr$ (squares) as
         a function of distance from the center of Abell 400.
         South lies to the left at negative distances, north to
         the right. The overall shapes of both PAs show
         a flatter region towards south, and a steepening towards
         north.}
\end{figure}

\begin{figure}
\centering
\resizebox{\hsize}{!}{\includegraphics{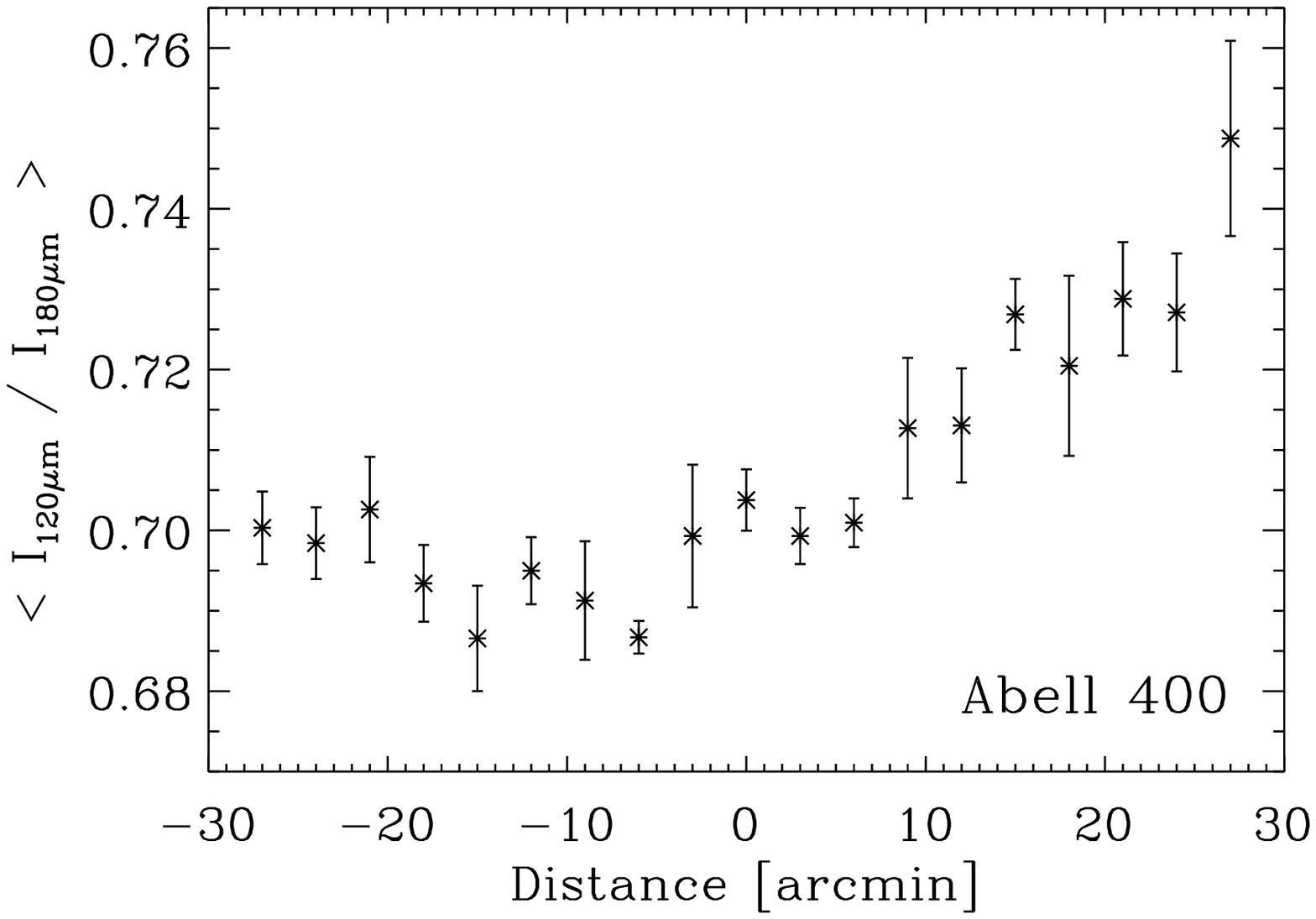}}
\caption[h3052f10.ps]{\label{fig:Abell400Overall}
         The overall zodiacal-light subtracted surface brightness ratio
         $\mathrm{I_{120 \mu m} / I_{180 \mu m}}$ for Abell 400,
         averaged over both position angles and all detector pixels.
         }
\end{figure}

\par
A dynamical analysis of Abell 400 was carried out by \citet{Beersetal92},
who concluded that this spiral rich cluster
actually consists of two bound sub-clusters currently undergoing
merging.  The irregular X-ray morphology \citep{Beersetal92,BuoteTsai96}
and the absence of a cooling flow \citep{Whiteetal97,AllenFabian97}
also indicates a young dynamical cluster age.  The overall X-ray temperature
of $\approx$ 2.5 keV is rather low \citep{Davidetal93,Whiteetal97}.

\par
Although Abell 400 lies at a much higher galactic latitude than
e.g.  Abell 262, both ISOPHOT scans (Fig. \ref{fig:Abell400measured})
show nevertheless a much higher absolute level of the 120\,$\mathrm{\mu m}$
surface brightness. The 100\,$\mathrm{\mu m}$
IRAS HIRES image shows quite a structured region with
large patches of cirrus foreground towards Abell 400.  Two very bright
point sources were detected with one detector pixel only along PA
40$\degr$ and PA 130 $\degr$, which have been identified with PGC
11141 (IRAS F02542+0600) and UGC 2444 (IRAS 02558+0606), respectively.
Both scans also indicate in the data streams of several pixels the presence
of a much weaker compact source confined to the sky position
centered on the cluster.

\par
The $\mathrm{I_{120 \mu m} / I_{180 \mu m}}$ raw surface brightness
ratios (Fig. \ref{fig:Abell400RatiosWZodi}) along the two PAs are
quite structured on angular scales of 10$\arcmin$ -- 20$\arcmin$
without a good overall agreement of the two profiles.  However, the
ratio at the scan crossing is in almost perfect agreement, thereby
providing an independent check of the flux calibration.  After
subtraction of the zodiacal light, the $\mathrm{I_{120 \mu m} / I_{180 \mu m}}$
surface brightness ratios (Fig. \ref{fig:Abell400Ratios})
of the two PAs are in much better agreement, showing a
flatter part towards south and a gradual steepening towards north.
This overall behavior is quite easily seen
in the overall $\mathrm{I_{120 \mu m} / I_{180 \mu m}}$ surface
brightness ratio (Fig. \ref{fig:Abell400Overall}).

\subsection {Abell 496}

\begin{figure}
  \centering
  \resizebox{\hsize}{!}{\includegraphics{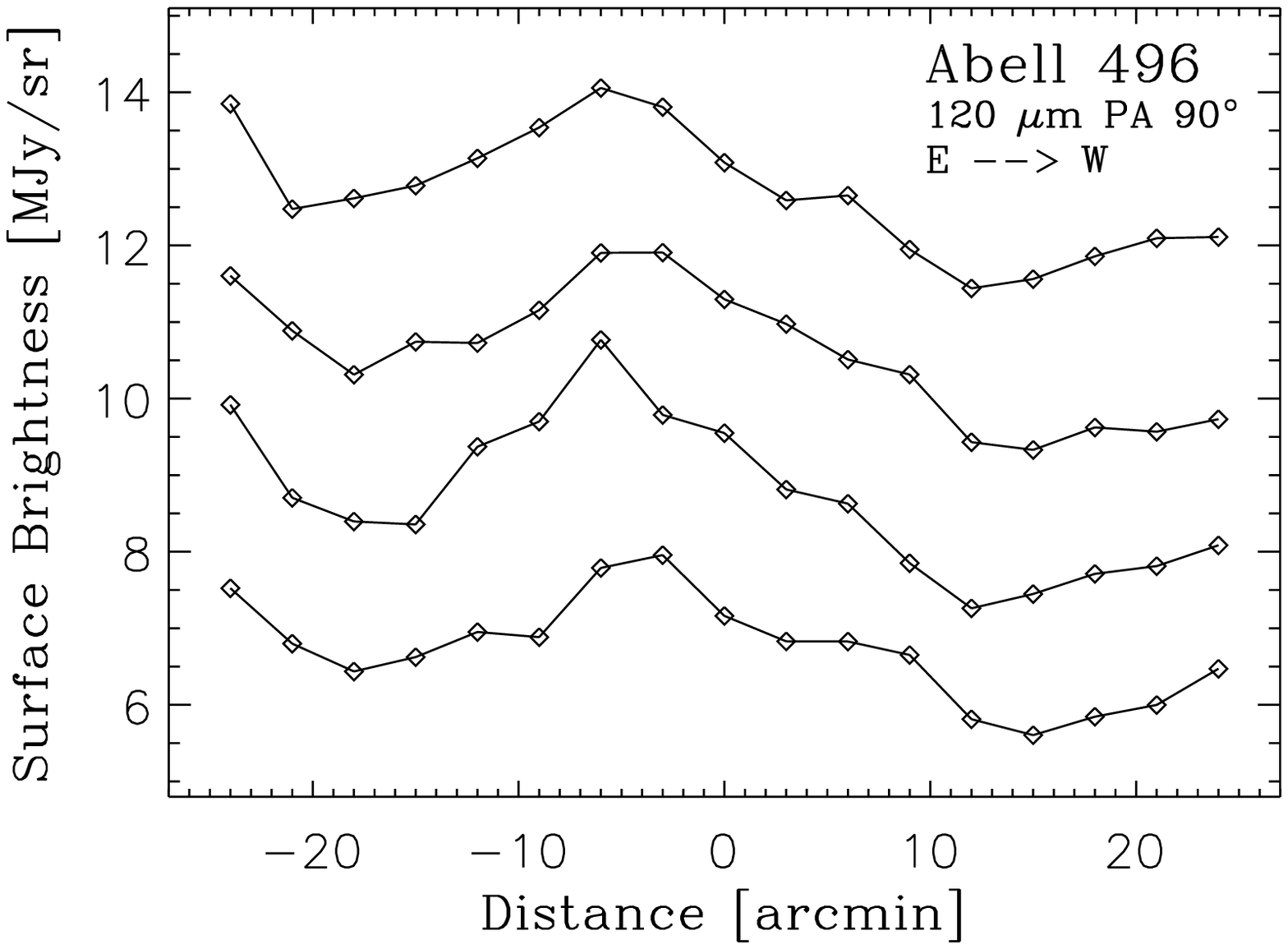}}
  \resizebox{\hsize}{!}{\includegraphics{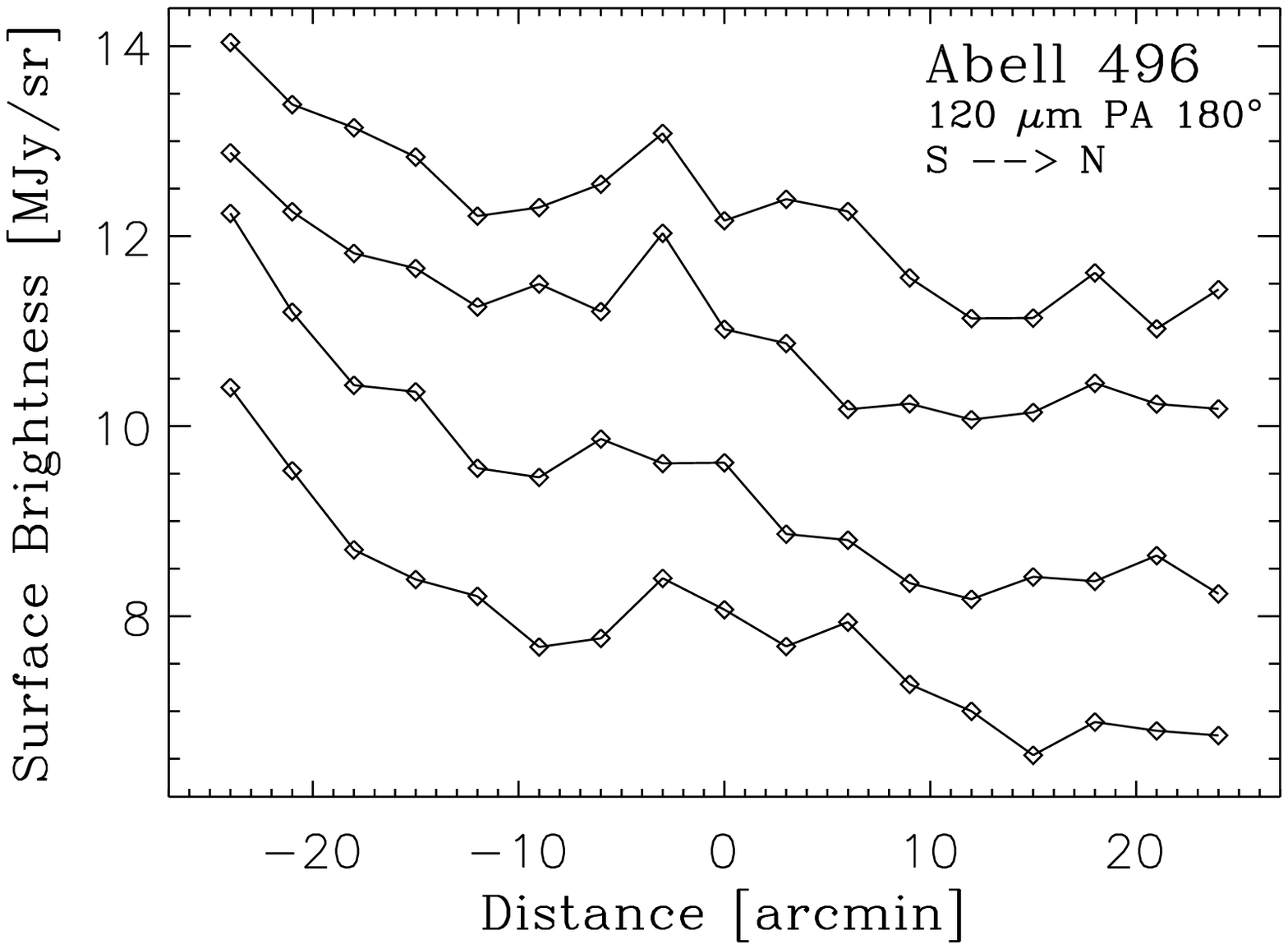}}
  \caption[]{\label{fig:Abell496measured}
           The observed brightness distributions of the four C200 detector
           pixels at 120\,$\mathrm{\mu m}$ for Abell 496 along PA
           90$\degr$ (top) and PA 180$\degr$ (bottom).  The brightness
           level is correct only for the lowest data stream. For
           clarity, the other three pixel data streams are offset
           arbitrarily. The 180\,$\mathrm{\mu m}$ brightness
           distributions (not shown) are quite similar.}
\end{figure}

\begin{figure}
\centering
\resizebox{\hsize}{!}{\includegraphics{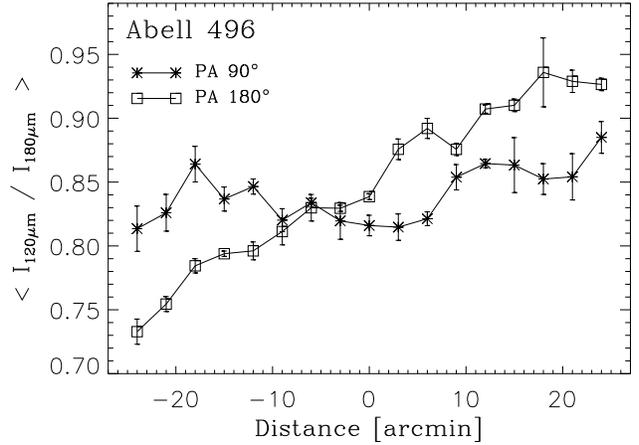}}
\caption[h3052f12.ps]{\label{fig:Abell496RatiosWZodi}
         The raw surface brightness ratios $\mathrm{I_{120 \mu m} / I_{180 \mu m}}$,
         averaged over all four detector pixels,
         along PA 90$\degr$ (asterisks, from east towards
         west) and PA 180$\degr$ (squares, from south towards north)
         as a function of distance from the center of Abell 496.
         The profile is nearly constant along PA 90$\degr$, while
         PA 180$\degr$ shows an almost linear increase.}
\end{figure}

\begin{figure}
\centering
\resizebox{\hsize}{!}{\includegraphics{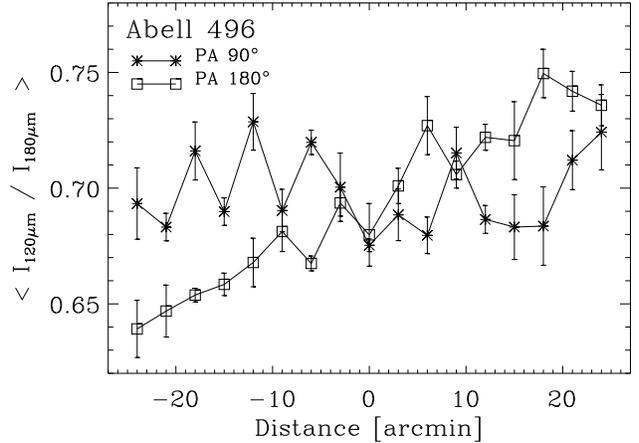}}
\caption[h3052f13.ps]{\label{fig:Abell496Ratios}
         The surface brightness ratios $\mathrm{I_{120 \mu m} / I_{180 \mu m}}$
         after subtraction of the Zodiacal light,
         averaged over all four detector pixels,
         along PA 90$\degr$ (asterisks, from east towards
         west) and PA 180$\degr$ (squares, from south towards north)
         as a function of distance from the center of Abell 496.}
\end{figure}

\begin{figure}
\centering
\resizebox{\hsize}{!}{\includegraphics{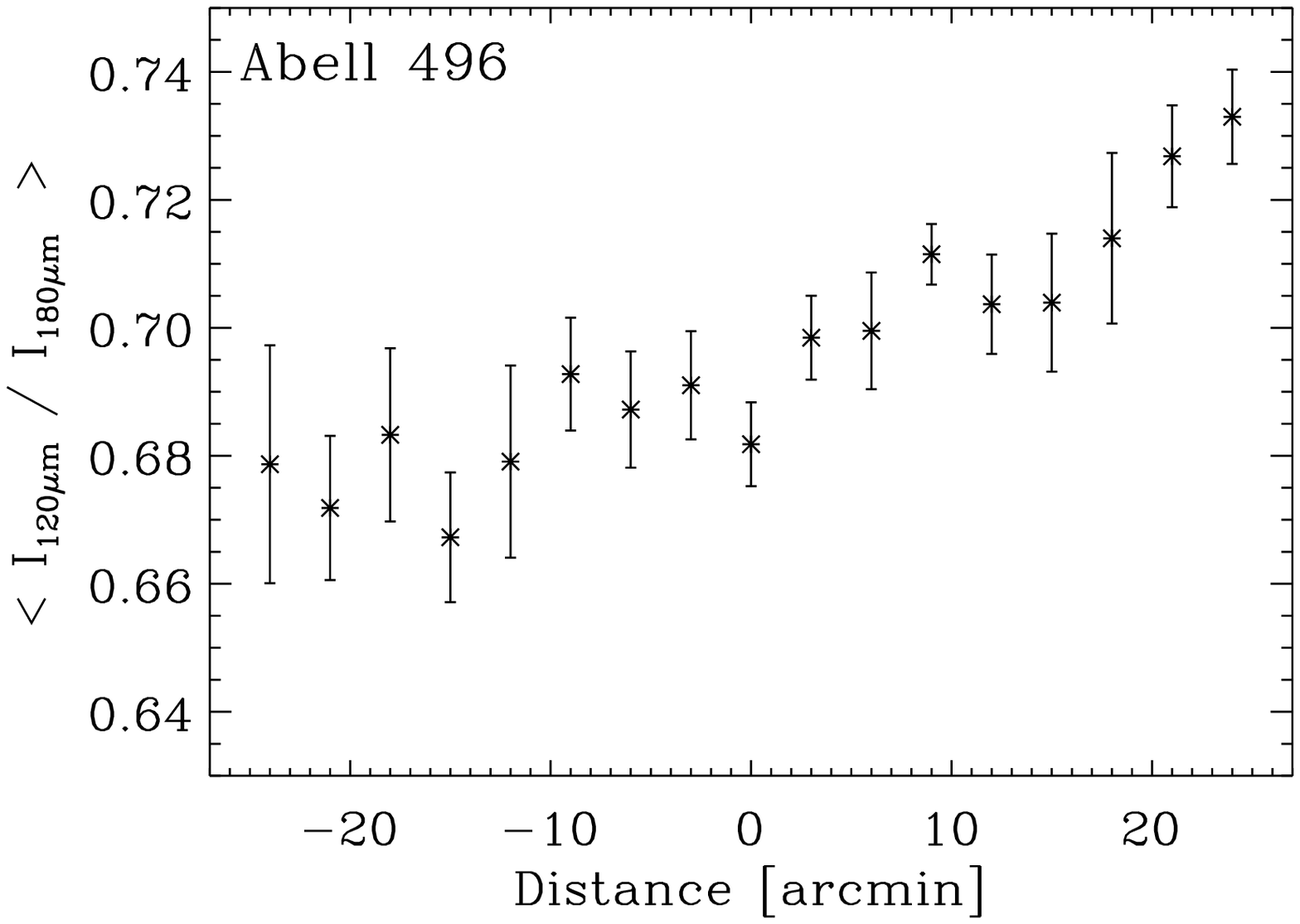}}
\caption[h3052f14.ps]{\label{fig:Abell496Overall}
         The overall zodiacal-light subtracted surface brightness ratio
         $\mathrm{I_{120 \mu m} / I_{180 \mu m}}$ for Abell 496,
         averaged over all four detector pixels.}
\end{figure}

\par
The relaxed single symmetric X-ray morphology \citep{BuoteTsai96},
the significant cooling flow (Tab. \ref{tab:physicalproperties}),
and the double-nucleus central cD galaxy with very small velocity
dispersion \citep{Tonry85} indicates a dynamically old, relaxed cluster.
There is evidence for an
X-ray absorption column density above the galactic value across
the cluster \citep{Whiteetal94,MacKenzieetal96,Allenetal01},
consistent with the excess
reddening of $\mathrm {E_{B-V} \approx}$ 0.20 mag derived from
emission line ratios \citep{Hu92}.

\par
The ISOPHOT scans along both PAs (Fig. \ref{fig:Abell496measured}) are
devoid of strong point sources. In the data streams of two pixels
along PA 180$\degr$, there is an indication for a weak compact source
confined to a single sky position.  A broad hump offset to the east
from the cluster center is visible along PA 90$\degr$, and a similar
but weaker bump appears to be superposed on
the generally decreasing surface brightness profiles along PA 180$\degr$.

\par
The two raw $\mathrm{I_{120 \mu m} / I_{180 \mu m}}$ surface
brightness ratios including the zodiacal light
(Fig. \ref{fig:Abell496RatiosWZodi}) along PA 180$\degr$ and PA
90$\degr$ are strikingly different. The former is almost monotonically
increasing from south towards north across the cluster, while the
latter varies much less across the cluster.  The color profiles after
subtraction of the zodiacal light (Fig. \ref{fig:Abell496Ratios}) do
not change very much, still showing a steady increase from south to
north and a rather flat but wiggling behavior along PA 90$\degr$.
Nevertheless, both profiles were combined to check for the
presence of structure on scales of 10$\arcmin$ -- 20$\arcmin$.
The resulting profile  (Fig. \ref{fig:Abell496Overall}), however,
shows only a gentle rise.

\subsection {Abell 1656 (Coma)}

\begin{figure}
  \centering
  \resizebox{\hsize}{!}{\includegraphics{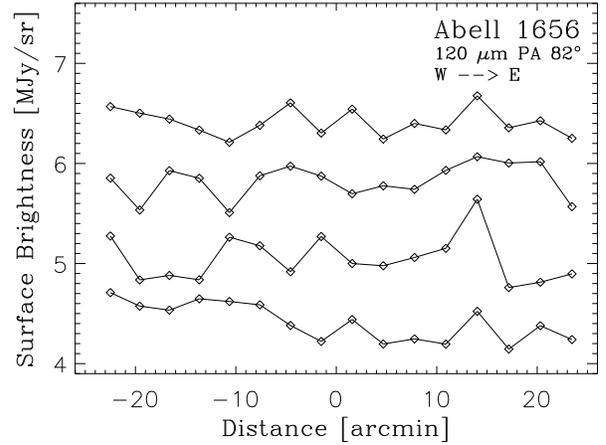}}
  \resizebox{\hsize}{!}{\includegraphics{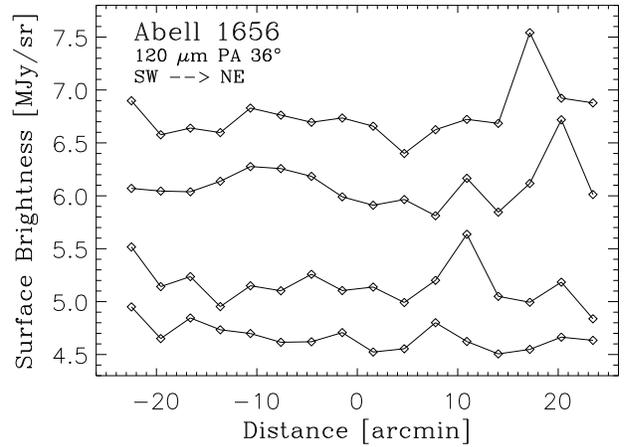}}
  \caption[]{\label{fig:Abell1656measured}
           The observed brightness distributions of the four C200 detector
           pixels at 120\,$\mathrm{\mu m}$ for Abell 1656 along PA
           82$\degr$ (top) and PA 36$\degr$ (bottom). The brightness
           level is correct only for the lowest data stream. For
           clarity, the other three pixel data streams are offset
           arbitrarily. The 180\,$\mathrm{\mu m}$ brightness
           distributions (not shown) are quite similar.}
\end{figure}

\begin{figure}
\centering
\resizebox{\hsize}{!}{\includegraphics{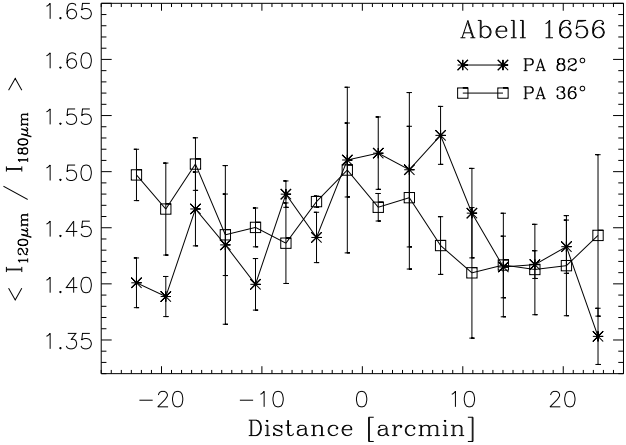}}
\caption[h3052f16.ps]{\label{fig:Abell1656RatiosWZodi}
         The raw surface brightness ratios $\mathrm{I_{120 \mu m} / I_{180 \mu m}}$,
         averaged over all four detector pixels,
         along PA 82$\degr$ (asterisks) and PA 36$\degr$ (squares) as
         a function of distance from the center of Abell 1656.  Both
         distributions show a broad bump closely aligned with the
         cluster center, indicating a cold
         component with $\mathrm{I_{120 \mu m}} > \mathrm{I_{180 \mu m}}$.}
\end{figure}

\begin{figure}
\centering
\resizebox{\hsize}{!}{\includegraphics{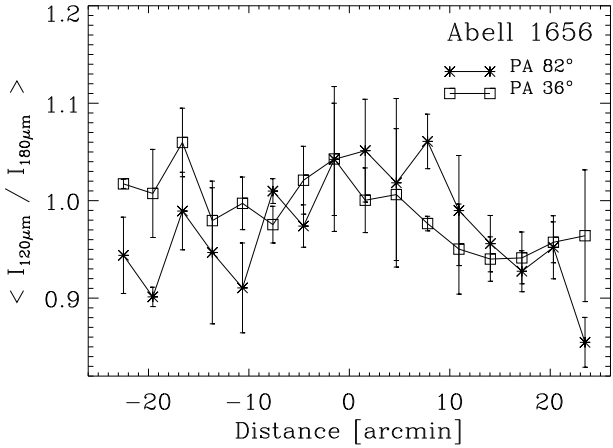}}
\caption[h3052f17.ps]{\label{fig:Abell1656Ratios}
         The surface brightness ratios $\mathrm{I_{120 \mu m} / I_{180 \mu m}}$
         after subtraction of the Zodiacal light,
         averaged over all four detector pixels,
         along PA 82$\degr$ (asterisks) and PA 36$\degr$ (squares) as
         a function of distance from the center of Abell 1656. The
         profiles are almost unchanged compared to
         Fig. \ref{fig:Abell1656RatiosWZodi}.}
\end{figure}

\begin{figure}
\centering
\resizebox{\hsize}{!}{\includegraphics{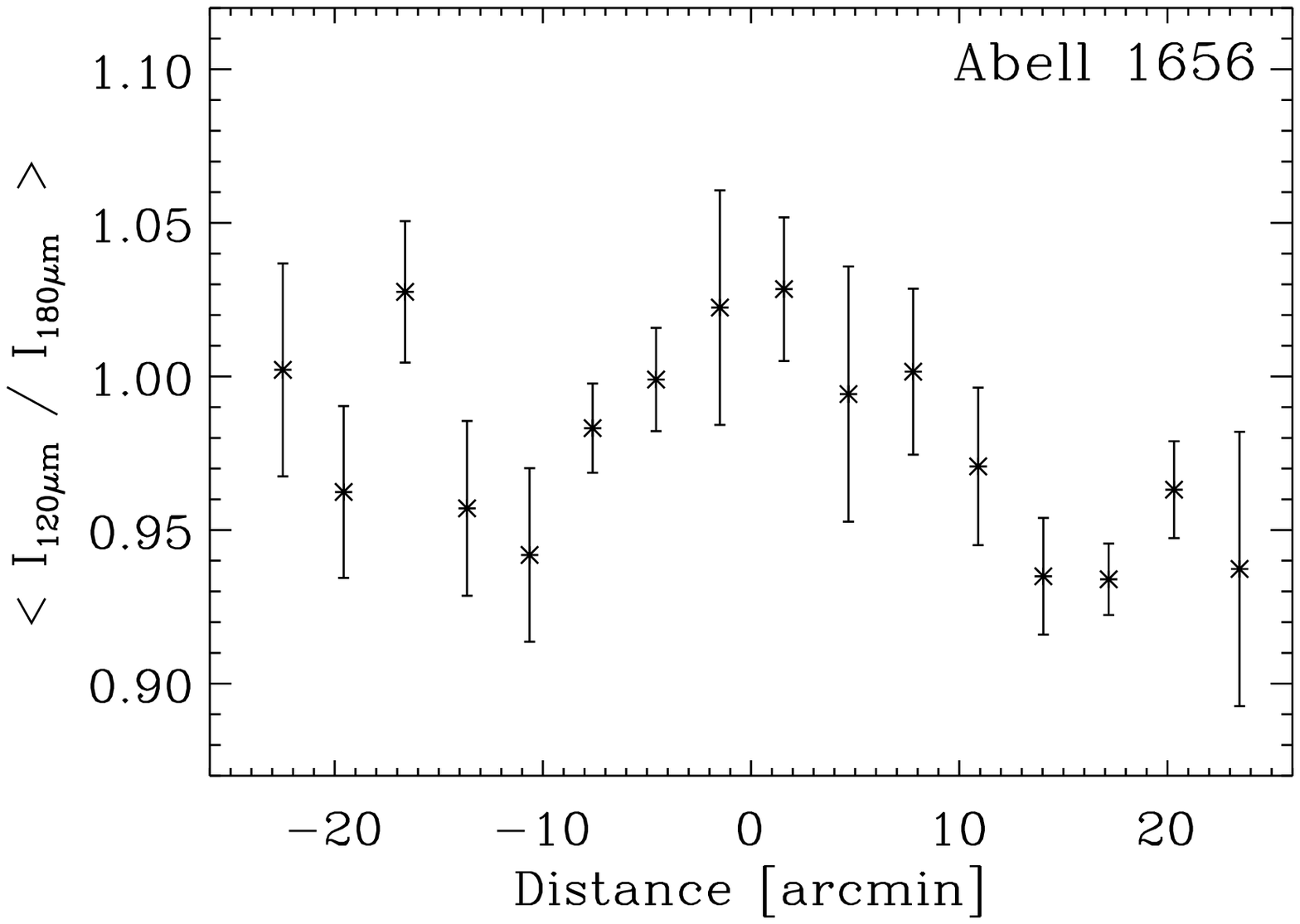}}
\caption[h3052f18.ps]{\label{fig:Abell1656Overall}
         The overall zodiacal-light subtracted surface brightness ratio
         $\mathrm{I_{120 \mu m} / I_{180 \mu m}}$ for Abell 1656
         averaged over both position angles and all detector pixels.}
\end{figure}

\par
There is now overwhelming evidence for recent or ongoing merging
processes in the Coma cluster from optical and X-ray data
\citep{CollessDunn96,Vikhlininetal97,Burnsetal94b,Neumannetal01}.  The
early evidence for intracluster light \citep{WelchSastry71,Mattila77} has
recently been confirmed with the detection of additional low surface
brightness structures interpreted as tidal debris from recent
galaxy interactions \citep{GreggWest98,TrenthamMobasher98}.  With an
X-ray temperature of $\approx$ 8 keV, Abell 1656 is the hottest of the
observed clusters, and in fact lies near the upper end of the X-ray
luminosity - temperature correlation \citep{Tuckeretal98}.

\par
Except for a slight decrease in surface brightness level due to the
updated ISOPHOT calibration, the scans along PA 82$\degr$ and PA 36$\degr$
(Fig. \ref{fig:Abell1656measured}) appear almost unchanged compared to
those shown in \citet{Stickeletal98}. However, due to the improved
data processing, there is now a much closer agreement between the
$\mathrm{I_{120 \mu m} / I_{180 \mu m}}$ surface brightness ratios
along the two PAs (Fig. \ref{fig:Abell1656RatiosWZodi}), showing significant
deviations only at the outer ends of the scans. An increased
$\mathrm{I_{120 \mu m} / I_{180 \mu m}}$ ratio within $\approx$
10$\arcmin$ from the cluster center is apparent, which
essentially remains unchanged even after the zodiacal light has been
subtracted (Fig. \ref{fig:Abell1656Ratios}).  The overall
$\mathrm{I_{120 \mu m} / I_{180 \mu m}}$ ratio averaged over both PAs
and all detector pixels (Fig. \ref{fig:Abell1656Overall}) shows the
bump quite clearly, which can be quite well represented by a gaussian
with a peak 5\% above a somewhat slanted background.

\begin{figure}
  \centering
  \resizebox{\hsize}{!}{\includegraphics{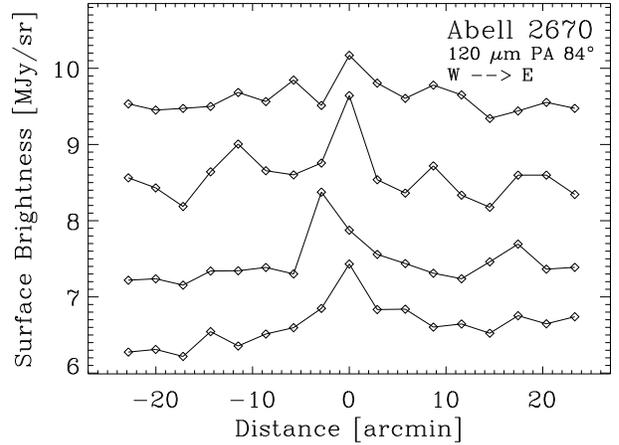}}
  \resizebox{\hsize}{!}{\includegraphics{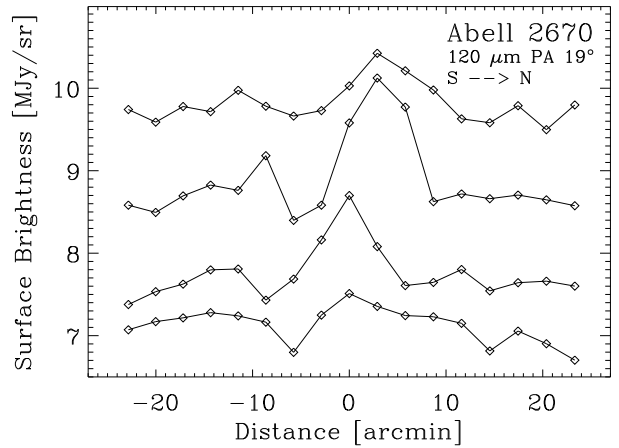}}
  \caption[]{\label{fig:Abell2670measured}
           The observed brightness distributions of the four C200 detector
           pixels at 120\,$\mathrm{\mu m}$ for Abell 2670 along PA
           84$\degr$ (top) and PA 19$\degr$ (bottom). The brightness
           level is correct only for the lowest data stream. For
           clarity, the other three pixel data streams are offset
           arbitrarily. The 180\,$\mathrm{\mu m}$ brightness
           distributions (not shown) are quite similar.}
\end{figure}

\begin{figure}
\centering
\resizebox{\hsize}{!}{\includegraphics{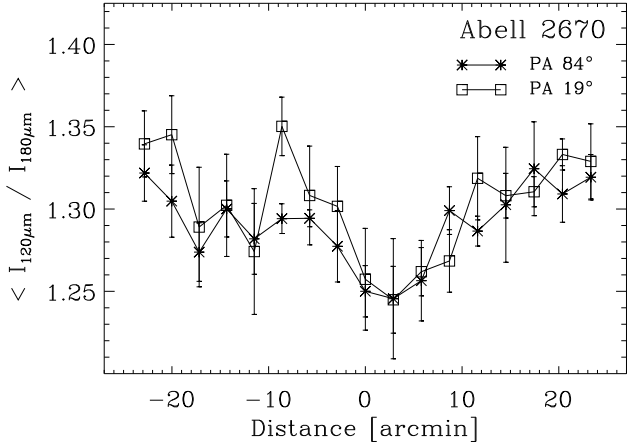}}
\caption[h3052f20.ps]{\label{fig:Abell2670RatiosWZodi}
         The raw surface brightness ratios $\mathrm{I_{120 \mu m} / I_{180 \mu m}}$,
         averaged over all four detector pixels,
         along PA 84$\degr$ (asterisks) and PA 19$\degr$ (squares) as
         a function of distance from the center of Abell 2670.}
\end{figure}

\begin{figure}
\centering
\resizebox{\hsize}{!}{\includegraphics{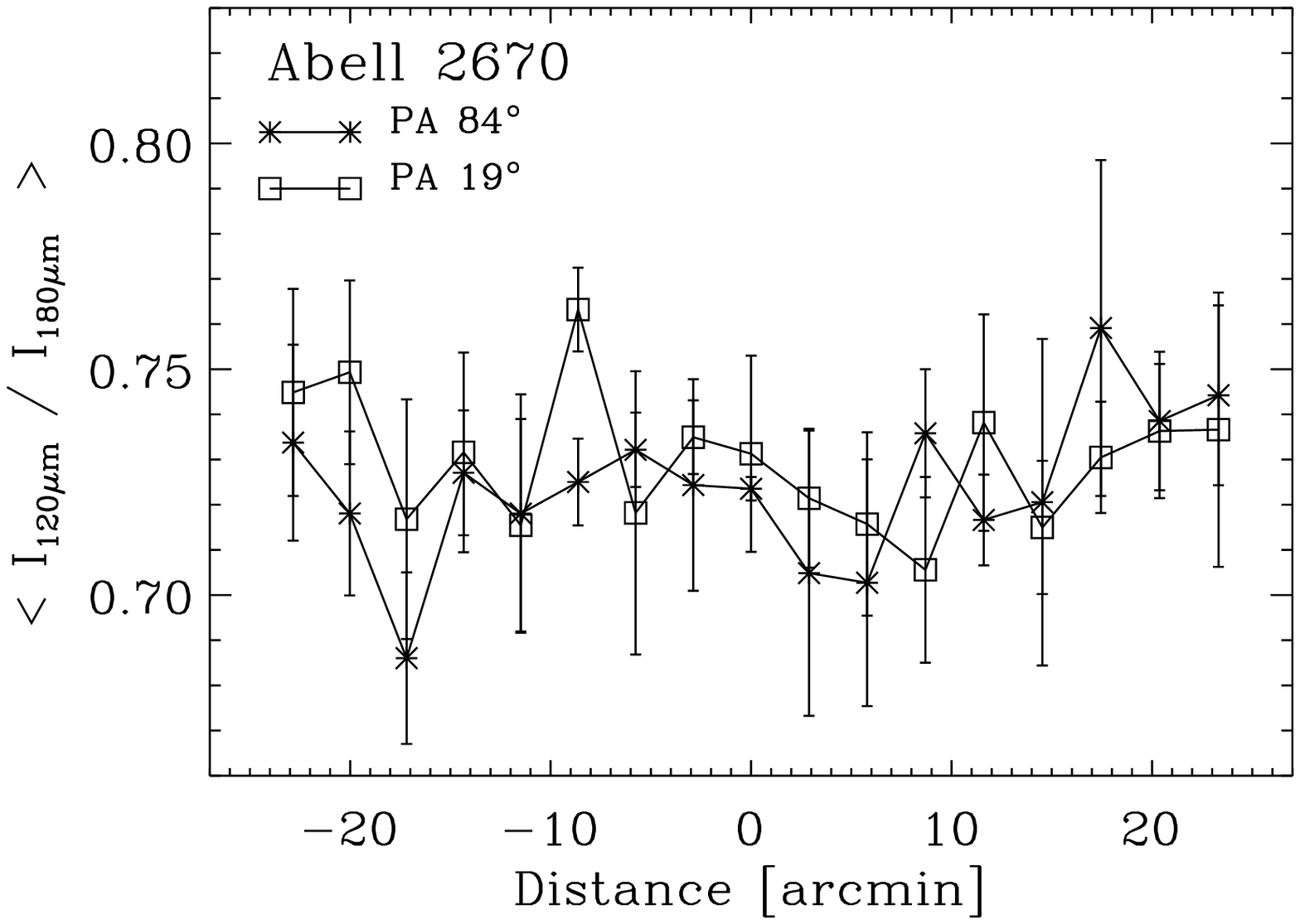}}
\caption[h3052f21.ps]{\label{fig:Abell2670Ratios}
         The surface brightness ratios $\mathrm{I_{120 \mu m} / I_{180 \mu m}}$
         after subtraction of the Zodiacal light,
         averaged over all four detector pixels,
         along PA 84$\degr$ (asterisks) and PA 19$\degr$ (squares) as
         a function of distance from the center of Abell 2670.}
\end{figure}

\begin{figure}
\centering
\resizebox{\hsize}{!}{\includegraphics{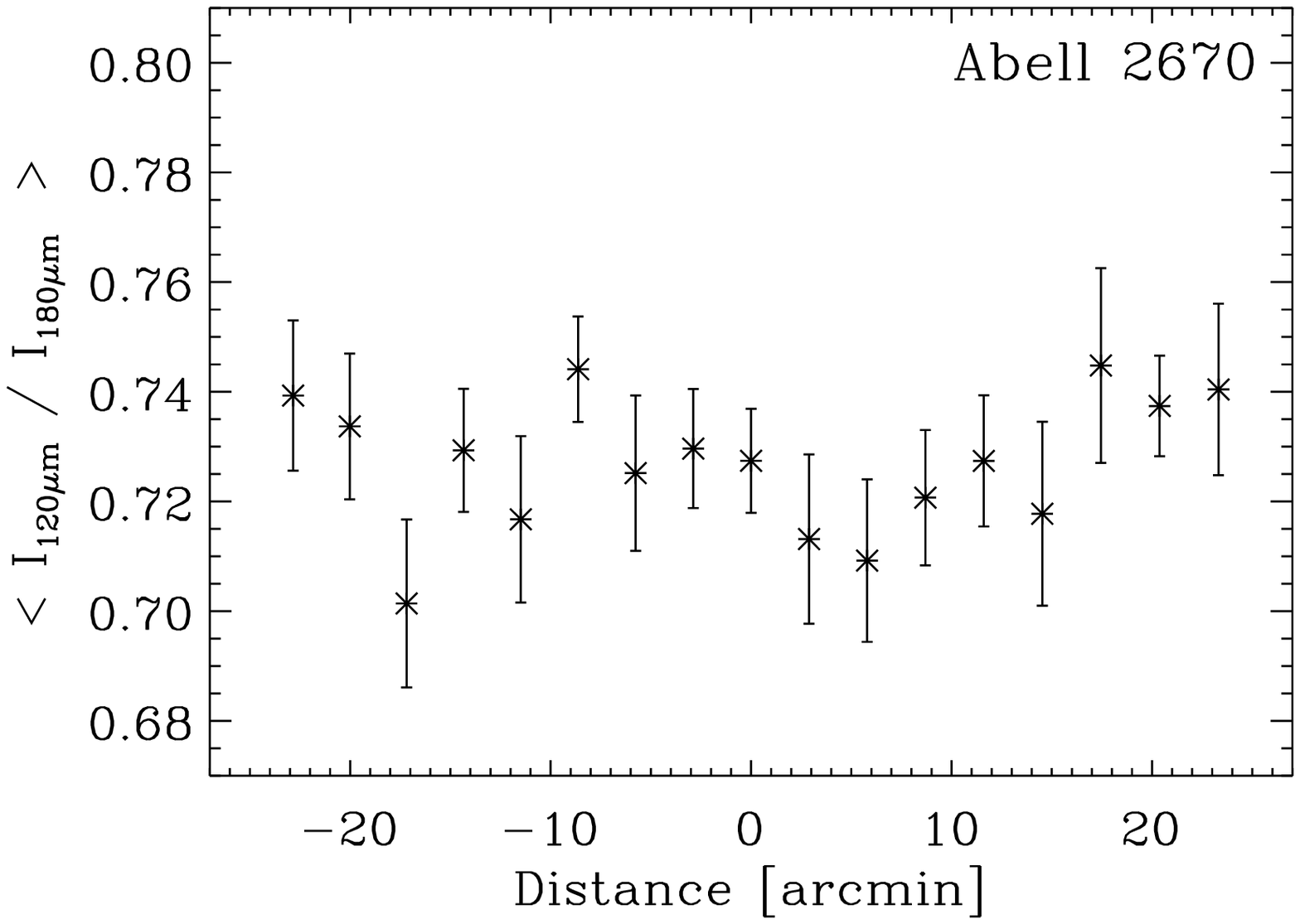}}
\caption[h3052f22.ps]{\label{fig:Abell2670Overall}
         The overall zodiacal-light subtracted surface brightness ratio
         $\mathrm{I_{120 \mu m} / I_{180 \mu m}}$ for Abell 2670
         averaged over both position angles and all detector pixels.}
\end{figure}

\subsection {Abell 2670}

\par
The X-ray morphology of Abell 2670 has a double peak
in the center \citep{BuoteTsai96},
where the secondary peak is due to several point sources, and
a pronounced asymmetry at low flux levels almost along the
north-south direction \citep{HobbsWillmore97}.  The ISOPHOT scan
position angles were chosen to cover both features.

\par
There are conflicting results regarding the dynamical stage of Abell
2670.  \citet{Bird94}, \citet{Escaleraetal94}, and
\citet{HobbsWillmore97} argue in favour of significant substructure
as a result of merging, while \citet{Sharplessetal88} and
\citet{Girardietal95} pointed out the contamination by a foreground
cluster.  This controversial picture is reinforced by the analysis of
X-ray data, where conflicting results about a possible cooling flow
have been found (see Tab. \ref{tab:physicalproperties}).

\par
Extended FIR emission was described by \citet{Wiseetal93}, but
judged as uncertain due to the relatively large cluster redshift and small
angular extent, and also by \citet{Coxetal95}. This diffuse emission is
easily seen on the 100 $\mathrm{\mu m}$ IRAS HiRes image. However,
rather than being centered on the cluster, it covers only the
western part of the cluster with a north-south elongated patch of low
level flux, overlaid with some slightly resolved brighter structures.
The central component of these structures is seen in the middle of the
overall rather flat surface brightness distributions along both
PA 84$\degr$ and PA 19$\degr$ (Fig. \ref{fig:Abell2670measured}).

\par
The $\mathrm{I_{120 \mu m} / I_{180 \mu m}}$ surface brightness ratios
(Fig. \ref{fig:Abell2670RatiosWZodi}) from both PAs show a general
similarity, a narrow central dip closely aligned with the cluster
centre, increasing to approximately the same level at the outer
edges of the scans. After subtraction of the zodiacal light,
the central dip vanishes in both individual
scans (Fig. \ref{fig:Abell2670Ratios}), resulting in an almost flat
overall  $\mathrm{I_{120 \mu m} / I_{180 \mu m}}$ ratio across the cluster
with only small scale low-level variations (Fig. \ref{fig:Abell2670Overall}).

\begin{figure}
  \centering
  \resizebox{\hsize}{!}{\includegraphics{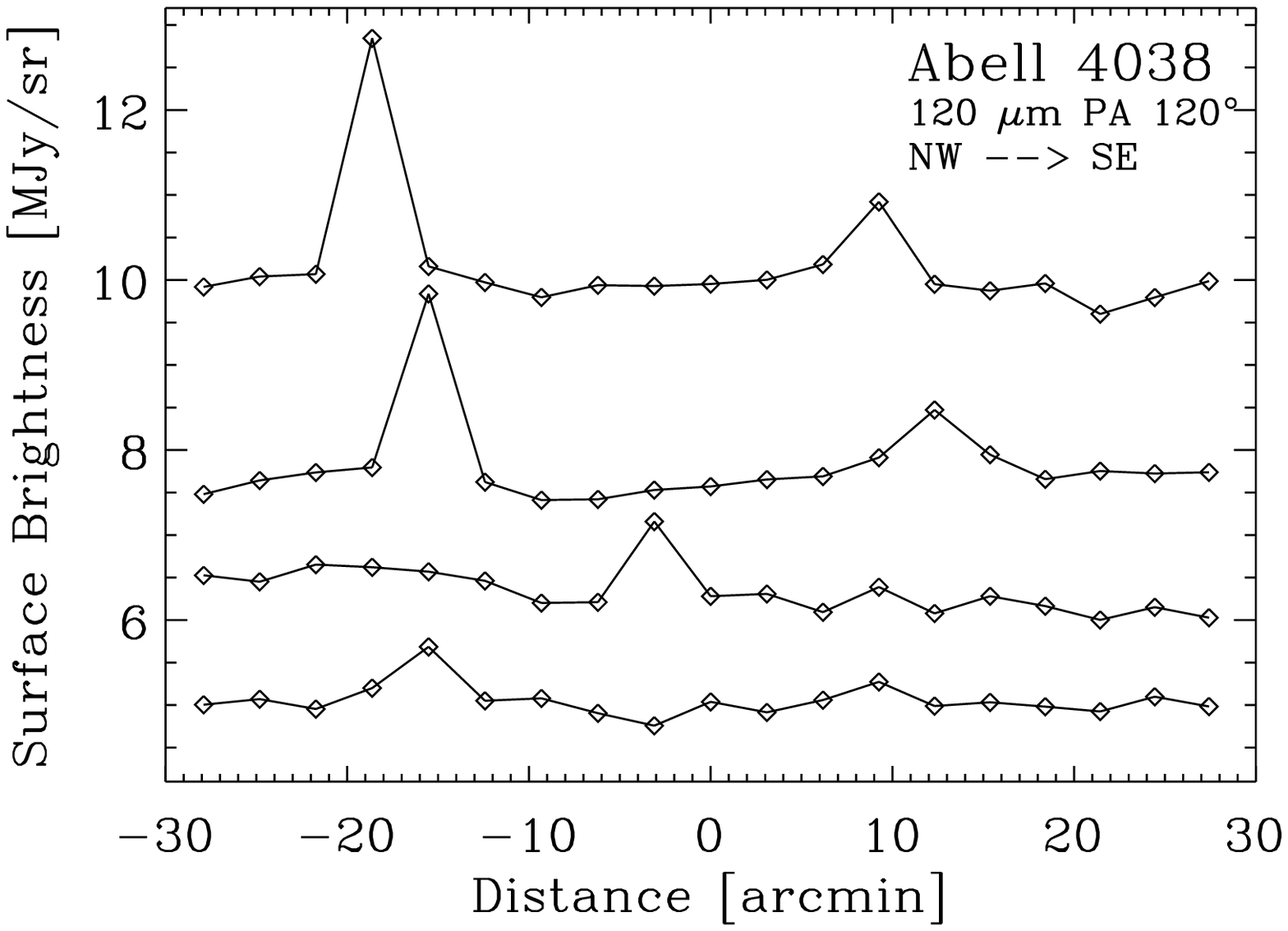}}
  \resizebox{\hsize}{!}{\includegraphics{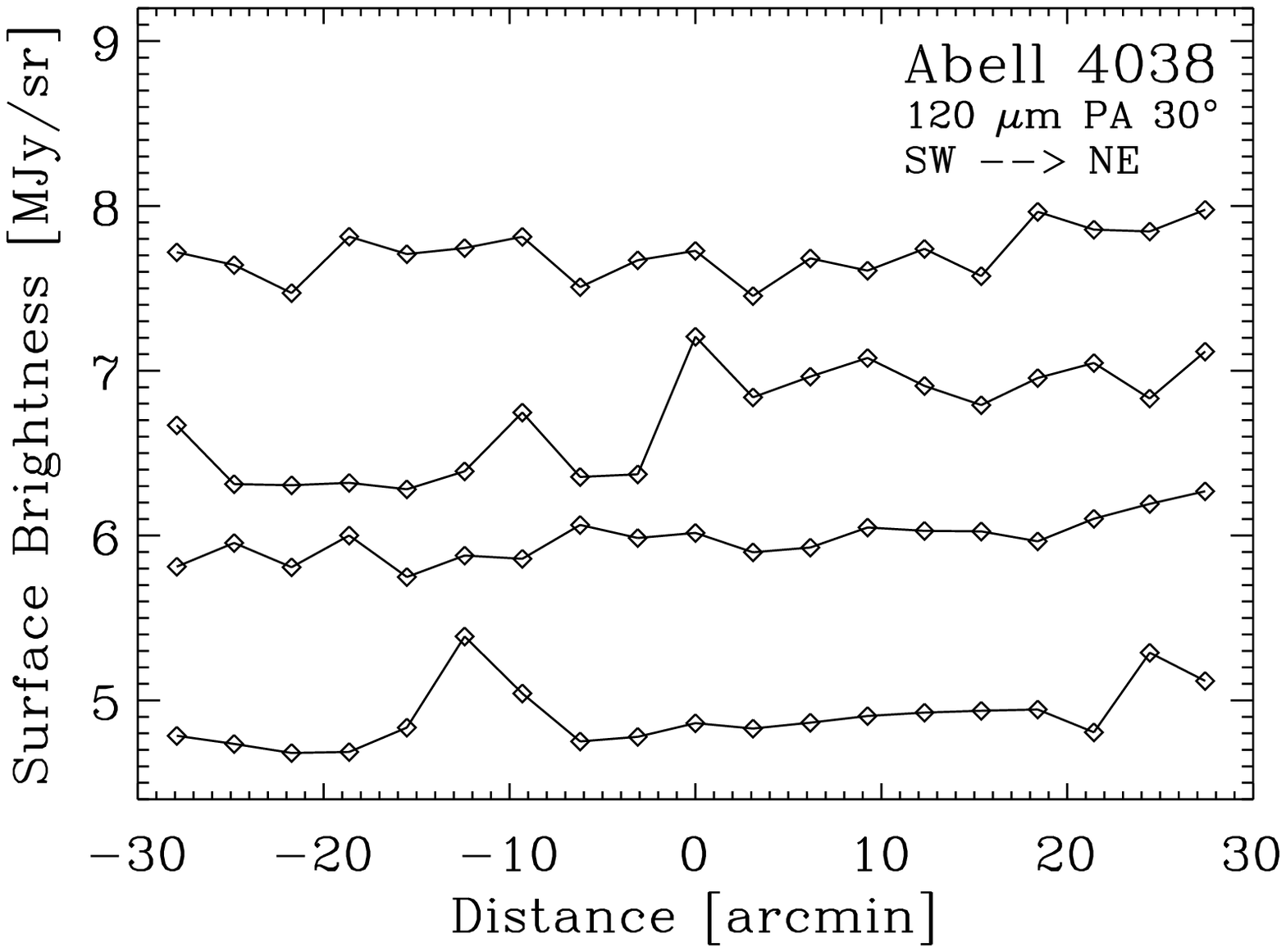}}
  \caption[]{\label{fig:Abell4038measured}
           The observed brightness distributions of the four C200 detector
           pixels at 120\,$\mathrm{\mu m}$ for Abell 4038 along PA
           120$\degr$ (top) and PA 30$\degr$ (bottom). The brightness
           level is correct only for the lowest data stream. For
           clarity, the other three pixel data streams are offset
           arbitrarily. The 180\,$\mathrm{\mu m}$ brightness
           distributions (not shown) are quite similar.\\
           In the scan along PA 30$\degr$ (bottom), one pixel data
           stream shows a marked jump in surface brightness near the
           cluster center, not seen in any
           other pixel or wavelength. This is most likely an abrupt change
           in pixel sensitivity not accounted for by the calibration.
           This pixel data stream has therefore been ignored in the subsequent
           analysis.}
\end{figure}

\begin{figure}
\centering
\resizebox{\hsize}{!}{\includegraphics{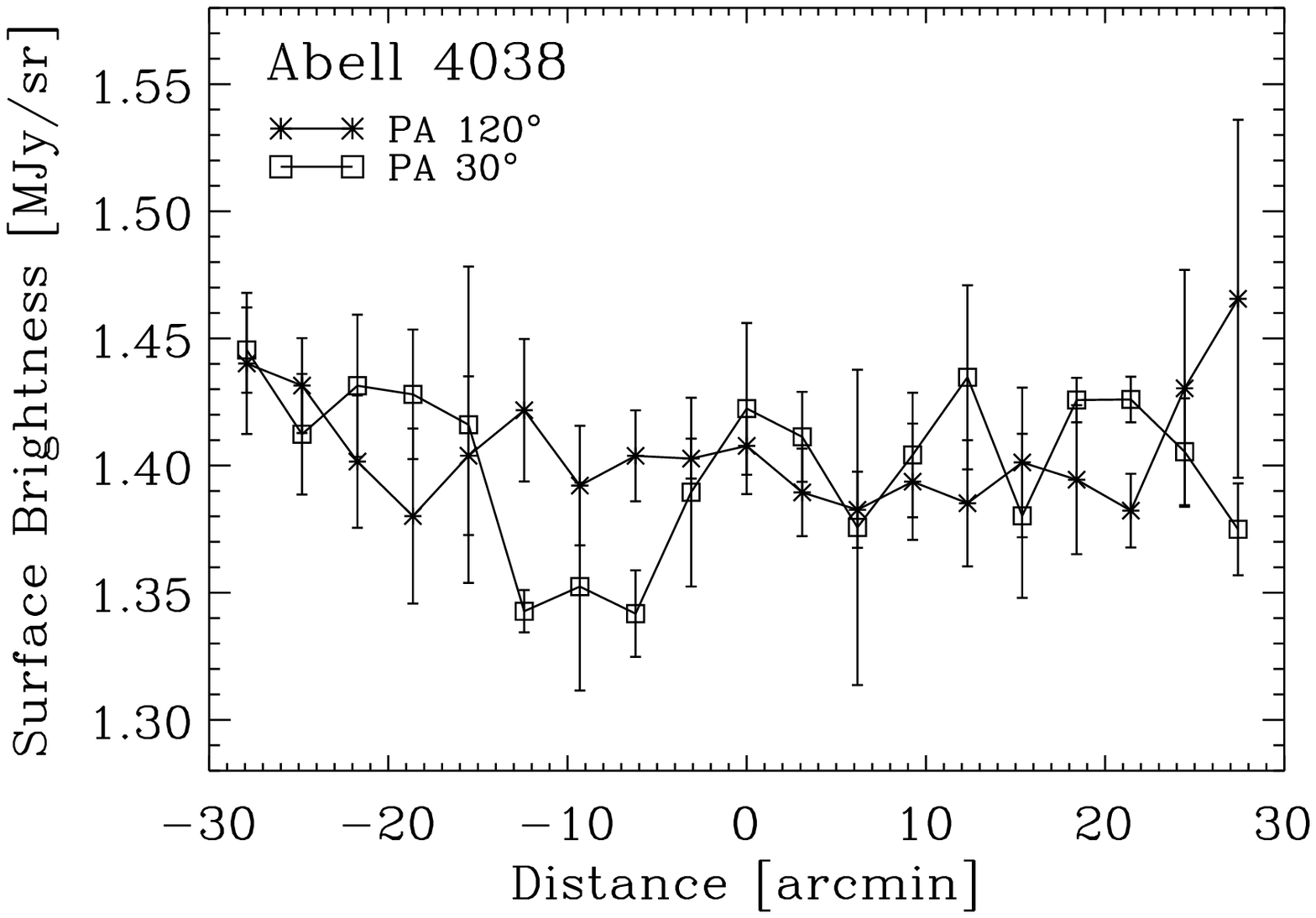}}
\caption[h3052f24.ps]{\label{fig:Abell4038RatiosWZodi}
         The raw surface brightness ratios $\mathrm{I_{120 \mu m} / I_{180 \mu m}}$,
         averaged over all four detector pixels,
         along PA 120$\degr$ (asterisks, average of four pixels) and
         PA 30$\degr$ (squares, average of three pixels) as
         a function of distance from the center of Abell 4038.}
\end{figure}

\begin{figure}
\centering
\resizebox{\hsize}{!}{\includegraphics{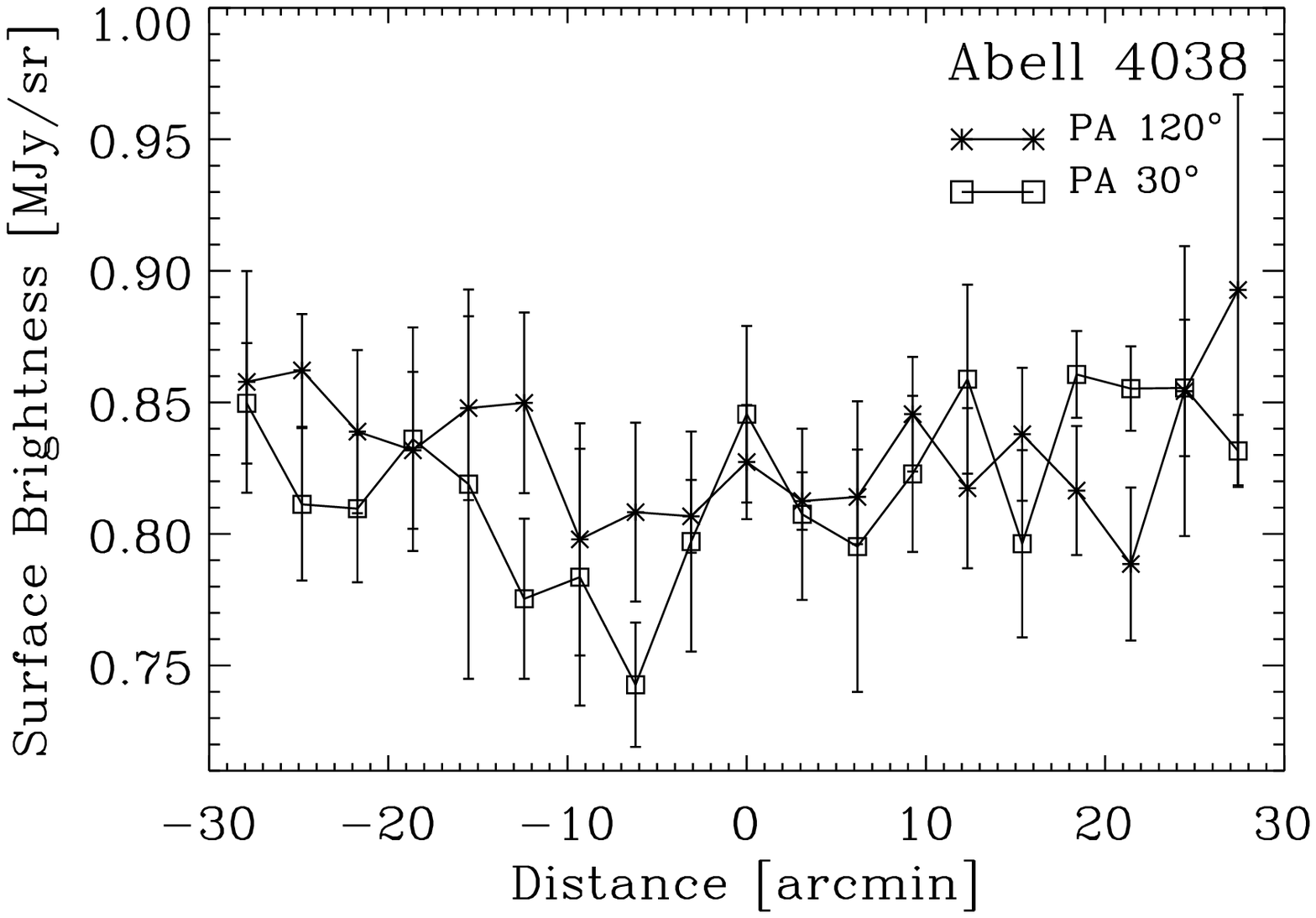}}
\caption[h3052f25.ps]{\label{fig:Abell4038Ratios}
         The surface brightness ratios $\mathrm{I_{120 \mu m} / I_{180 \mu m}}$
         after subtraction of the Zodiacal light,
         averaged over all four detector pixels,
         along PA 120$\degr$ (asterisks, average of four pixels) and
         PA 30$\degr$ (squares, average of three pixels) as
         a function of distance from the center of Abell 4038.}
\end{figure}

\begin{figure}
\centering
\resizebox{\hsize}{!}{\includegraphics{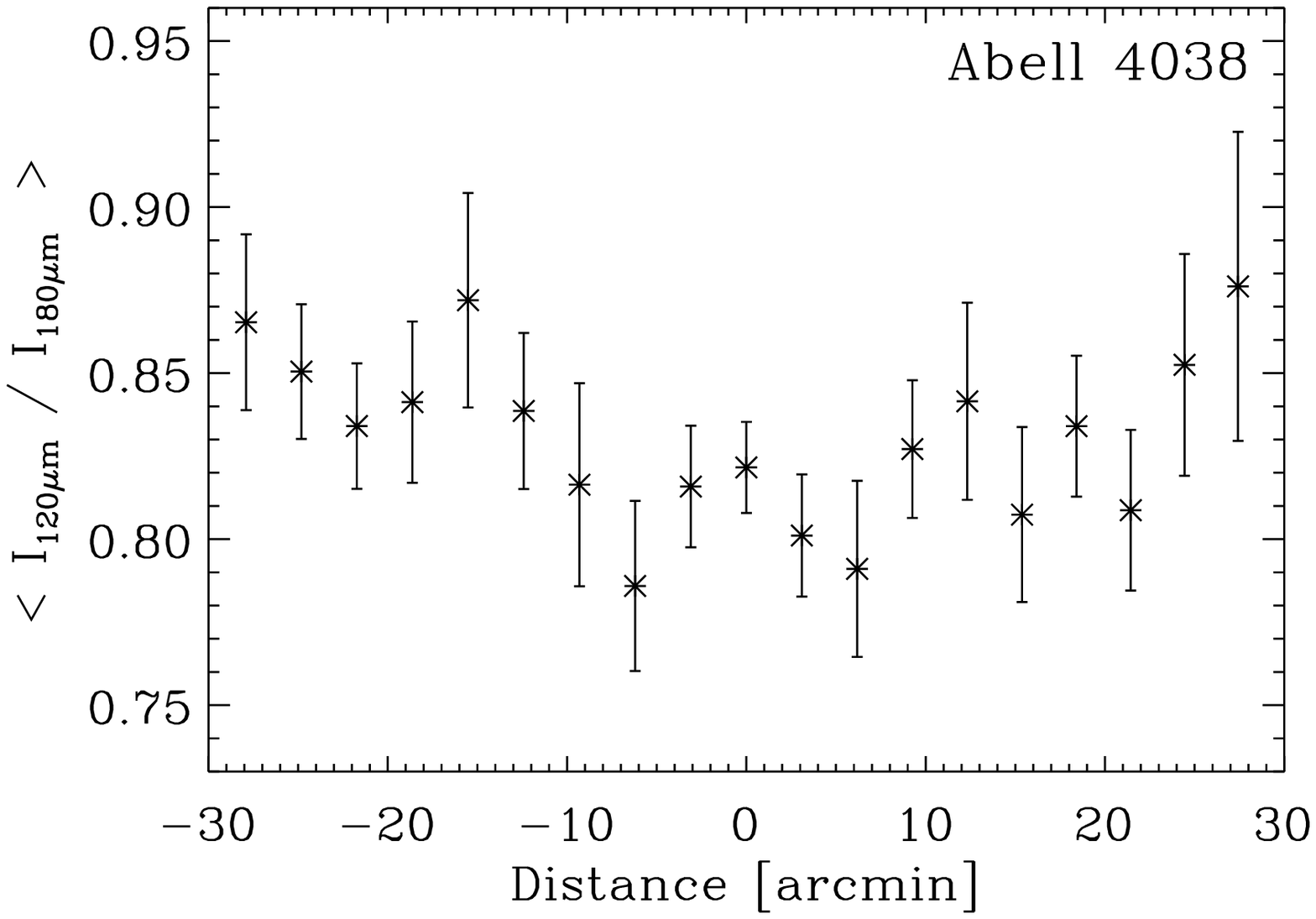}}
\caption[h3052f26.ps]{\label{fig:Abell4038Overall}
         The overall zodiacal-light subtracted surface brightness ratio
         $\mathrm{I_{120 \mu m} / I_{180 \mu m}}$ for Abell 4038
         averaged over both position angles and all detector
         pixels.}
\end{figure}

\subsection {Abell 4038 (Klemola 44)}

\par
The X-ray morphology of Abell 4038 is similar to that of the Coma
cluster, being essentially elliptical with a faint extended source
representing a group of galaxies within the cluster \citep{BuoteTsai96}.
The X-ray data indicate the
presence of a cooling flow (see Tab. \ref{tab:physicalproperties}).
The overall galaxy distribution is elongated along PA 120$\degr$, as
is the major axis of the central cD galaxy.  \citep{Greenetal90}. One
of the ISOPHOT scans was aligned with this direction while the other
one was perpendicular.

\par
The overall surface brightness distributions along both PA 120$\degr$
and PA 30$\degr$ (Fig. \ref{fig:Abell4038measured}) are rather
flat. Only along PA 120$\degr$ compact sources show up in the data
streams of single detector pixels.  The brightest source detected along
PA 120$\degr$ is actually a background galaxy, identified with IRAS
23440-2816 (z = 0.0495).

\par
The $\mathrm{I_{120 \mu m} / I_{180 \mu m}}$ surface brightness ratios
(Fig. \ref{fig:Abell4038RatiosWZodi}) along the two PAs agree well, being
generally flat with only small-scale  structures. The same behavior
is seen after zodiacal-light subtraction (Fig. \ref{fig:Abell4038Ratios})
and eventually in the overall $\mathrm{I_{120 \mu m} / I_{180 \mu m}}$
profile (Fig. \ref{fig:Abell4038Overall}).

\section {Discussion}

\par
The striking result of the analysis of
the $\mathrm{I_{120 \mu m} / I_{180 \mu m}}$ zodiacal-light subtracted
surface brightness ratios is the presence of an easily noticeable bump
in only one cluster, namely Abell 1656 (Coma), the cluster with the
highest X-ray temperature and no indication of a cooling flow.

\par
Although the raw $\mathrm{I_{120 \mu m} / I_{180 \mu m}}$
profiles showed structures on scales of 10$\arcmin$ -- 20$\arcmin$
also for Abell 262 and Abell 2670, these structures
disappeared with the subtraction of the zodiacal light
component.  This indicates that an additional FIR emitting component
outside the Galaxy is present in the direction of Abell 1656 while
galactic cirrus foreground emission close to the line-of-sight is
responsible for the structures in the color profiles of Abell 262 and
Abell 2670. This is in clear contrast to the findings of \citet{Wiseetal93},
who ascribed residual FIR emission towards Abell
262 and Abell 2670 found on IRAS ISSA plates to ICD. However, in these
two as well as the other three observed clusters, the presence of ICD
cannot be excluded completely, because it might be present at very
low levels or have properties similar to galactic cirrus.
In the former case, the signature is easily lost in the noise,
while in the latter case there is no way of discriminating it against
the foreground galactic dust even with photometric measurements at
additional FIR wavelengths.

\par
The bump in the $\mathrm{I_{120 \mu m} / I_{180 \mu m}}$ ratio in
Abell 1656 (Coma) might be caused by the integrated FIR
emission of the unresolved cluster galaxies, if their ISM had different
temperatures and/or dust properties than the foreground galactic
cirrus \citep{Quillenetal99}.  Unless the galaxy content of Abell 1656
were rather special, the other clusters should
also show a characteristic signature in
their $\mathrm{I_{120 \mu m} / I_{180 \mu m}}$ profiles,
which, however, is not observed. Moreover, the FIR color temperature
determination of a large sample of normal galaxies selected from the
ISOPHOT Serendipity Survey \citep{Stickeletal00} showed the ubiquitous
presence of dust in the ISM with temperatures of T $\approx$ 17\,K
similar to the Galaxy.

\par
In principle, the Sunyaev-Zeldovich effect might be another
cluster intrinsic cause for extended excess emission at FIR and sub-mm
wavelengths. However, even for very hot clusters, the effect at the
peak wavelength of $\approx \mathrm{800 \mu m}$ is small
\citep[e.g.][]{Komatsuetal99}, and vanishes shortward of
$\approx \mathrm{300 \mu m}$ \citep{Chaseetal87,Rephaeli95}, making it
very unlikely as a cause for the observed systematic changes in the
$\mathrm{I_{120 \mu m} / I_{180 \mu m}}$ profiles.

\par
The most likely interpretation of the observed spatially extended
bump in the $\mathrm{I_{120 \mu m} / I_{180 \mu m}}$ ratio of Abell
1656 is therefore the presence of intracluster dust
mixed with the hot X-ray emitting electron gas. The excess surface brightness
indicates that the intracluster dust has properties different from the
foreground galactic cirrus such as temperature, grain sizes, grain
composition, or emissivity, otherwise no signature in the
$\mathrm{I_{120 \mu m} / I_{180 \mu m}}$ ratios would be discernible.

\par
The observed overall zodiacal light subtracted
$\mathrm{I_{120 \mu m} / I_{180 \mu m}}$ ratio $R^{obs}$
can be written as

\begin{equation}
R^{obs} = \frac{I^{cirr}_{120} + I^{ICD}_{120}}
               {I^{cirr}_{180} + I^{ICD}_{180}}~,
\label{equ:relation4}
\end{equation}

\noindent
where $\mathrm{I^{cirr}_{120}}$ and $\mathrm{I^{ICD}_{120}}$ are the surface
brightnesses of the foreground galactic cirrus and the excess emission
of the cluster at 120\,$\mathrm{\mu m}$, respectively, and
$\mathrm{I^{cirr}_{180}}$ and $\mathrm{I^{ICD}_{180}}$ the
corresponding brightnesses at
180\,$\mathrm{\mu m}$.
To quantify the bump in the overall zodiacal-light subtracted
$\mathrm{I_{120 \mu m} / I_{180 \mu m}}$ surface brightness ratio
of Abell 1656, it was modelled as the sum of a gaussian, presumed to
result from the cluster dust emission $\mathrm{I^{ICD}}$,
and a first order polynomial from the large scale cirrus
foreground $\mathrm{I^{cirr}}$. The cirrus foreground can then be
interpolated at the position of the cluster centre, giving the
central $\mathrm{I^{cirr}_{C120} / I^{cirr}_{180}}$ ratio

\begin{equation}
R^{cirr} = \frac{I^{cirr}_{120}}{I^{cirr}_{180}}
\label{equ:relation5}
\end{equation}

\noindent
of the cirrus alone.

\par
For the case of $\mathrm{R^{obs} > R^{cirr}}$,
the two relations (Eqs. \ref{equ:relation4},\ref{equ:relation5})
can be re-arranged to give

\begin{equation}
I^{ICD}_{120} = I^{cirr}_{180}\,\times\,\left[(R^{obs} - R^{cirr}) +
                         R^{obs}\,\frac{I^{ICD}_{180}}{I^{cirr}_{180}} \right].
\end{equation}

\noindent
Under the assumption  that the cluster dust emission is weak compared to the
cirrus foreground, i.e. $\mathrm{I^{ICD}_{180} \ll I^{cirr}_{180}}$, it
can further be rearranged to

\begin{equation}
\label{equ:dustexcess120}
I^{ICD}_{120} \simeq ({I^{cirr}_{180} + I^{ICD}_{180}}) \,\times\, (R^{obs} - R^{cirr}).
\end{equation}

\noindent
This gives the cluster excess surface brightness
$\mathrm{I^{ICD}_{120}}$ in terms of the observed overall
$\mathrm{I_{120 \mu m} / I_{180 \mu m}}$ ratio $\mathrm{R^{obs}}$, the
interpolated $\mathrm{I_{120 \mu m} / I_{180 \mu m}}$ ratio
$\mathrm{R^{cirr}}$ of the cirrus, and the observed total surface
brightness at 180\,$\mathrm{\mu m}$.

For Abell 1656, $\mathrm{R^{obs}-R^{cirr}}$ = 0.07, with a
characteristic e-folding angular scale of  $\mathrm{\sigma_{ICD}} \approx 5\arcmin$,
and $\mathrm{I^{cirr}_{180} + I^{ICD}_{180}}$ = 3.1 MJy/sr, resulting in a
cluster excess peak surface brightness $\mathrm{I^{ICD}_{120}}$ $\approx$ 0.2 MJy/sr.
Assuming that the fitted one-dimensional gaussian is representative of
an underlying circular symmetric two-dimensional excess flux
distribution, the integrated (total) excess fluxes computed
from $\mathrm{F_{ICD} = 2\,\pi\,\sigma_{ICD}^2\,I^{ICD}}$, the volume of the
two-dimensional gaussian with a peak height equal to the derived
excess surface brightness and a width equal to the fitted
one-dimensional gaussian, is $\mathrm{F_{ICD}} \approx$  2.8 Jy.
This corresponds to a monochromatic FIR luminosity
$\mathrm{L_{ICD} = 4\,\pi\,D^{2}\nu\,F_{ICD}}$ of
$\mathrm{L_{ICD} \approx 8.6\times10^{43}erg/s}$, where
D is the cluster luminosity distance.

\par
The absence of a clear sign of cluster excess emission in the
individual scans of Abell 1656 at both wavelengths gives strict upper
limits for the ICD excess surface brightness
of $\mathrm{I^{ICD}_{120}} < 0.4$\,MJy/sr.  Under the assumption that
the profile of the ICD excess emission at both wavelengths can be
approximated by a gaussian and the cirrus foreground surface
brightness varies linearly across the cluster, this absolute upper
limit can be transformed to a lower limit of
for the peak surface brightness
ratio $\mathrm{I^{ICD}_{120} / I^{ICD}_{180}}$
of the excess emission. This is done by fitting the
ratio of the presumed surface brightness profiles (for each wavelength
a gaussian added to a slanted background) to the overall
$\mathrm{I_{120 \mu m} / I_{180 \mu m}}$ ratio, subject to the
constraint not to exceed the upper limit, yielding
$\mathrm{I^{ICD}_{120} / I^{ICD}_{180}} \ga  2$.
This lower limit in the $\mathrm{I^{ICD}_{120} / I^{ICD}_{180}}$ ratio
transforms into a lower limit in the dust color temperature of
$\mathrm{T \ga 30\,K}$, assuming a blackbody (Planck) function
modified by a frequency dependent emissivity
$\mathrm{F_{\nu} = {\nu}^\mathrm{\beta} B_{\nu}(T_\mathrm{D})}$
with exponent $\mathrm{\beta = 2}$.

\par
An estimation of the intracluster dust mass can be derived by using
\begin{equation}
  M_{D} = D^{2}\,F_{\nu}\,\left[\kappa_{\lambda}\,B_{\nu}(T_{D})\,\right]^{-1}
\end{equation}
\noindent
\citep{Hildebrand83,Draine90}, where $\mathrm{F_\nu}$ is the total
excess flux density at $\mathrm{120 \mu m}$, $\mathrm{T_{D}}$ the
dust color temperature, and
$\mathrm{\kappa_\lambda}$ the dust opacity.  With a dust opacity of 3
$\mathrm{m^{2}/kg}$, which lies in the middle of the range of
currently considered values \citep{Draine90}, a dust
temperature of $\mathrm{T_{D}} \approx$ 30 K and the excess flux of
$\mathrm{F_{ICD}}$ = 2.8 Jy a dust mass
of $\mathrm{M_{D} \approx 10^{7} \, M_{\sun}}$ is derived.  This dust
mass is only exemplary, since it depends strongly on the assumed dust
color temperature and the dust opacity.

\par
It should be noted that a dust opacity, which already
included the conversion to gas mass with a constant galactic
gas-to-dust ratio, had been used in the first
publication of the Coma observations \citep{Stickeletal98}.  The newly
derived dust mass for Abell 1656 is therefore considerably lower than
that given before. Consequently, the gas-to-dust ratio using a gas mass of
$\mathrm{M_{gas} \approx 10^{13}\,M_{\sun} }$ is extremely high
($\mathrm{\approx 10^{6}}$). Furthermore, the low inferred dust mass
leads under standard assumptions about the dust properties
\citep{Whittek92} to a negligible amount of visual extinction
($\mathrm{A_{V} \ll 0.1 mag}$), in
contrast to the reported optical extinction of $\approx$ 0.3 mag
\citep{Zwicky62,KarachentsevLipovetskii69}, and in accord with the
analysis of the IRAS data by \citet{Dweketal90}.

\section {Conclusions}

Spatially extended FIR excess emission from dust distributed in the
intracluster medium has been searched for in a small sample of six
Abell clusters covering a wide variety in cluster properties, such
as X-ray morphology and temperature, cluster age, richness, and
distance, by using ISOPHOT scan measurements at $\mathrm{120 \mu m}$
and $\mathrm{180 \mu m}$ wavelength.  After subtraction of the
zodiacal light, the characteristic signature of a dip or bump on
angular scales of 10$\arcmin$ -- 20$\arcmin$ is only seen in
Abell 1656 (Coma), a hot, dynamically young cluster with optical and
X-ray signs of interactions with smaller galaxy groups.

\par
The evidence from IRAS ISSA plates for FIR emission from intracluster
dust in Abell 262 and Abell 2670 \citep{Wiseetal93} could not be
confirmed. Although the raw (including zodiacal light)
$\mathrm{I_{120 \mu m} / I_{180 \mu m}}$ profiles did show a dip, it
disappeared after the zodiacal light had been subtracted.  This
behavior is indicative of a localized patch of galactic foreground
cirrus, which happens to lie along the line of sight to these
clusters. This strikingly underlines the necessity of interpreting FIR
measurements, even flux ratios at several wavelengths, carefully
by taking into consideration the ever-present cirrus structures
in conjunction with the zodiacal light, which can produce a quite
counterintuitive behavior.

\par
This result does seem to support the simple picture put forward in
\citet{Stickeletal98}, that only young clusters currently undergoing
merging show the signature of intracluster dust, which might indicate
that intracluster dust is primarily brought into the cluster from the
outside, either by merging with external galaxy groups or possibly
by steady infall of intercluster material \citep{Popescuetal00}.

\par
The FIR observations of the six clusters, particularly those with a
strong cooling flow inferred from X-ray observations, do not seem to
support the picture of cold clouds dropping out of the cooling flow
and shielding intracluster dust from sputtering, or even dust
originating in the dense centers of cooling flow clusters.  The rather
low inferred dust mass in Abell 1656 and the non-detection of the
characteristic signature of ICD in the other five clusters can be
taken as evidence that the X-ray excess absorption seen in cooling
flow clusters is likely not due to the presence of intracluster dust.
Particularly, no excess X-ray absorption is required for Abell 1656
\citep{Allenetal01}, the only cluster where evidence for extended FIR
emission has been found in this study.  The absence of a signature for
in\-tra\-cluster dust in the clusters with strong indication for cooling
flows is in agreement with several other unsuccessful searches for the
final stages of the cooled material, and adds additional doubt on the
correctness of material condensations associated with the cooling flow
model.

\par
Finally, intracluster dust appears not be responsible for the
the observed dimming of the high - redshift supernovae,
and the contribution to the FIR extragalactic background light
is likely also vanishingly small.

\begin{acknowledgements}
The development and operation of ISOPHOT were supported by
MPIA and funds from Deutsches Zentrum f\"ur Luft- und
Raumfahrt (DLR, formerly DARA). The ISOPHOT Data Centre
at MPIA is supported by Deutsches Zentrum f\"ur Luft- und
Raumfahrt (DLR) with funds of Bundesministerium f\"ur
Bildung und Forschung, grant. no. 50\,QI9801\,3.
The authors are responsible for the content of this publication. \\
This research has made use of the Digitized Sky Survey, produced at
the Space Telescope Science Institute, NASA's Astro\-physics Data
System Abstract Service, the Simbad Database, operated at CDS,
Strasbourg, France, and the NASA/IPAC Extragalactic Database (NED)
which is operated by the Jet Propulsion Laboratory, California
Institute of Technology, under contract with the National Aeronautics
and Space Administration. \\
We thank  Eli Dwek and Rick Arendt for useful comments on a draft of
the manuscript.
\end{acknowledgements}

\end  {document}